\documentclass[a4paper,11pt]{article}
\pdfoutput=1 

\usepackage{jheppub}

\usepackage[ddmmyyyy,hhmmss]{datetime}

\usepackage{amsmath,graphicx,amssymb,subfigure,bbm,comment}
\usepackage[all]{xy}
\usepackage{mathtools}
\usepackage{amsfonts}
\usepackage{xcolor}
\usepackage{autobreak}
\allowdisplaybreaks

\title{Holographic three-point correlators at finite density and temperature}

\author[a]{George Georgiou}
\author[a,b]{and Dimitrios Zoakos}

\affiliation[a]{Department of Physics, National and Kapodistrian University of Athens, 15784 Athens, Greece}
\affiliation[b]{Department of Engineering and Informatics, 
Hellenic American University, 436 Amherst st, Nashua, NH 03063 USA}

\emailAdd{ggeo@phys.uoa.gr}
\emailAdd{zoakos@gmail.com}

\abstract{We calculate holographically three-point functions of scalar operators with large dimensions at finite density and finite temperature. To achieve this, we construct new solutions that involve two isometries of the deformed internal space. 
The novel feature of these solutions is that the corresponding two-point function depends not only on the conformal dimension but also on the difference between the two angular momenta. After identifying the dual operators, we systematically calculate three-point correlators  as an expansion in powers of the temperature and the chemical potential. Our analytic perturbative results are in agreement with the exact numerical computation.  
The three point correlator (when
the background contains either temperature or density but not both) is always
a monotonic function of the temperature or the chemical potential. 
However, when
both parameters are present the three point correlator is no longer a monotonic function. 
For fixed finite
temperature and small values of the chemical potential a minimum of the three-point function appears. Surprisingly, contributions from the internal space do not depend on the chemical potential or the temperature, as long as those are treated as perturbations.}

\begin{document}
\maketitle
\flushbottom

\section{Introduction and background}
\label{intro}

Gauge/gravity dualities \cite{Maldacena:1997re,Gubser:1998bc,Witten:1998qj} provide an invaluable mean of addressing difficult questions in strongly coupled  gauge theories. In particular, the study of the strong coupling regime of conformal field theories (CFTs) at finite temperature and finite density (chemical potential) can be achieved by considering the near horizon limit of a stack of non-extremal rotating D3-branes \cite{Kraus:1998hv,Cvetic:1996dt}.

The main focus of this work will be on the holographic calculation of two and three-point correlation functions involving scalar operators with large conformal dimensions in the aforementioned framework. 
For such operators, the bulk-to-boundary propagator of the dual super-gravity field can be related, in the WKB approximation, to the geodesic length of the trajectory for a point particle traveling from the boundary point, where the field theory operator is inserted, up to a point in the bulk. Consequently, the computation of correlators boils down to the determination of certain geodesics.
Despite the fact that such a framework looks oversimplifying, the computation of the two and three-point correlation functions reveals a very interesting and rich structure and 
allows one to probe the physical properties of the dual CFT at strong coupling and in the presence of both finite temperature and finite density, or equivalently finite chemical potential. 

Significant effort has been put so far in the holographic calculation of two-point functions both in the absence (see for example \cite{Tsuji:2006zn,Georgiou:2010an,Georgiou:2011qk}) and in the presence 
of temperature \cite{Balasubramanian:1999zv,Louko:2000tp}.\footnote{For the holographic calculation in backgrounds that are not asymptotically $AdS$ see \cite{Fuertes:2009ex,Georgiou:2018zkt}.} 
Thermal two-point correlation functions contain more information than their zero temperature counterparts, 
due to their intricate structure. When the distance between the operators is much smaller than the size 
of the thermal circle, the operator product expansion (OPE) can be utilised. Moreover, operators will generically develop  a vacuum expectation value (VEV) while the presence of the thermal circle provides both a scale and a direction. 
The general structure of the thermal two-point functions has been discussed in \cite{Iliesiu:2018fao}. It was argued that two-point correlators
admit an expansion in terms of Gegenbauer polynomials. The microscopic 
data of the CFT, namely the structure constants of the OPE and the VEVs of operators enter in the coefficients of this expansion. Recently,  thermal two-point functions have been studied in the 
framework of holography in \cite{Rodriguez-Gomez:2021pfh, Rodriguez-Gomez:2021mkk} (see also \cite{Krishna:2021fus}). As expected, a systematic expansion of the two-point correlator in terms of the Gegenbauer polynomials was 
found. The coefficient corresponding  
to the energy-momentum tensor was compared successfully against preceding gravity computations. 

Geodesics in black hole backgrounds and their connection with thermal correlations functions is not a new topic.
It has been argued that thermal two-point functions may probe the interior of the black hole \cite{Fidkowski:2003nf, Festuccia:2005pi, Hubeny:2006yu, Hubeny:2012ry}. This idea has been revived recently 
in \cite{Grinberg:2020fdj} where it was argued that thermal 
one-point functions encode the proper time from the horizon to the black-hole singularity of a radially in-falling particle. 
This was implemented through the presence of higher curvature couplings (e.g. the square of the Weyl tensor $W^2$)  between the scalar field and the gravitons. Furthermore, one-point thermal functions in the limit of large dimensions and large spin have been considered in \cite{David:2022nfn,David:2023uya}. Subsequently, it was argued that a similar mechanism is at work, but now for the thermal two-point function \cite{Rodriguez-Gomez:2021pfh}.

In \cite{Georgiou:2022ekc},  the holographic study of one and two-point correlation functions using the geodesic 
approximation, was extended to field theories that are both at finite temperature and at finite R-charge (chemical potential).\footnote{The retarded Green’s functions in thermal CFTs with chemical potential and angular momenta was calculated in  \cite{Bhatta:2022wga,Bhatta:2023qcl}.}
One and two-point functions of scalar operators at finite density and/or finite temperature $T$ were calculated holographically. In the case of finite density and zero  temperature it was argued that only scalar operators can have non-zero VEVs. In the presence of  both  chemical potential and temperature, a systematic expansion of the two-point correlators in power of the temperature $T$ and the chemical potential $\Omega$ was presented.  It was found that the two-point function can be written as a linear combination of the Gegenbauer polynomials $C_J^{(1)}(\xi)$ but with the coefficients depending now on both the chemical potential and the temperature. The three leading terms in the OPE originate from the expectation values of the scalar operator $\phi^2$, the R-current ${\cal J}^\mu_{\phi_3}$ and the stress-energy tensor $T^{\mu\nu}$, respectively. Finally, by comparing the appropriate term of the holographic result for the two-point correlator to the corresponding term in the OPE, complete agreement was found between the value of the R-charge derived from the holographic result and the value of the R-charge obtained from the thermodynamics of the black hole.
The aim of this work is, on one hand to continue the study of two-point functions in the presence of both finite temperature and chemical potential, and on the other hand to proceed to the calculation of three point correlators.

In the rest of this section, we will review material necessary for the calculation of holographic correlators. 
From \cite{Georgiou:2022ekc}, we will need the expression of the gravity background that we will study, 
and more importantly the general method for computing two-point functions.   
The virtue of this method, that combines ideas from \cite{Janik:2010gc} and \cite{Rodriguez-Gomez:2021pfh}, is that it 
takes into account properly the motion of the particle in the internal space. 
We will extend this method to accommodate a more generic motion of the particle in the internal space, 
than the one that it was covered in \cite{Georgiou:2022ekc}.
Starting from the Polyakov action,  
the computation boils down to an effective action that can be used for calculating higher point correlation functions. 
More specifically, we are interested in the holographic computation of three-point functions. 
In the limit of operators with large conformal dimensions, the bulk-to-boundary propagator is approximated by the exponential of minus the mass times the geodesic length.
This geodesic corresponds to the trajectory of a particle that travels from the boundary point, 
where the operator is inserted, until the interaction point in the bulk.

The background that we will to study three-point correlation functions is the non-extremal rotating D3-brane solution of  
type-IIB supergravity. This solution was obtained in \cite{Russo:1998mm,Kraus:1998hv}  and its metric is
\begin{eqnarray}  \label{metric-general-v1}
ds^2 &=& H^{-\frac{1}{2}}\Bigg[-\left(1-\eta^4 \, H \right) dt^2 + d\vec{x}_3^2 \Bigg]  + H^{\frac{1}{2}}
\Bigg[\frac{dz^2}{f}  + \frac{\tilde \Delta}{z^2} \, d\theta^2 + \frac{1}{z^2} \, \cos^2\theta \, d\Omega_3^2 
\nonumber\\
&& + \, \frac{1}{z^2} \, \left(1-r_0^2 \, z^2\right) \sin^2\theta\ d\phi_1^2  
- 2 \, \eta^2 \, r_0 \,   \cos^2\theta  \, dt \left(\sin^2 \psi \, d\phi_2+ \cos^2 \psi \, d\phi_3\right) \Bigg] 
\end{eqnarray}
where the line element of the three-sphere is given by
\begin{equation}
d\Omega_3^2 \, = \, d\psi^2 + \sin^2 \psi \, d\phi_2^2+ \cos^2 \psi \, d\phi_3^2
\end{equation}
and the different functions $H$, $f$ and $\tilde \Delta $ of the metric are defined as follows
\begin{equation} \label{def-functions-general}
H  =  \frac{z^4}{\tilde \Delta}\ , \quad 
f =  H \, \left(1 - \eta^4 \, z^4 \right) \, \left(1 - r_0^2 \, z^2 \right)  \quad {\rm with} \quad 
\tilde \Delta = 1- r_0^2 \, z^2 \cos^2\theta\, .
\end{equation}
In the above expressions, $r_0$ denotes the angular momentum parameter and is related to the R-charge density (chemical potential) and $\eta$ determines the location of the horizon (i.e. $z_H = \eta^{-1}$) and is related to the temperature.
The precise thermodynamic relations are  the following \cite{Gubser:1998jb,Harmark:1999xt,Russo:1998by} 
\begin{equation} \label{thermo-quantities}
T= \frac{1}{\pi} \, \sqrt{\eta^2 - r_0^2} \quad \& \quad   
\Omega = r_0 \quad {\rm with} \quad   \eta \ge r_0 \, . 
\end{equation}

The analysis will focus on a consistent truncation of the ten-dimensional set of equations of motion. The ansatz for the point like string will be a 
generalization of one of the cases that were examined in \cite{Georgiou:2022ekc} which now takes into account the motion along two of the angles of the deformed 5-sphere, namely $\phi_2$ (with angular momentum $\omega_2$) 
and $\phi_3$  (with angular momentum $\omega_3$). 
For a point like string sitting at $\theta=0$ \& $\psi=\frac{1}{2}\arccos \psi_0$, the metric \eqref{metric-general-v1} takes the form
\begin{equation}  \label{metric-general-v2}
ds^2= \frac{1}{z^2} \Bigg[f_2 \, d\tau^2 +  
\sqrt{g} \, d\vec{x}_3^2 +\frac{dz^2 }{\sqrt{g}\, f_1} \Bigg] - \frac{1}{\sqrt{g}} \Big[ A \, d\phi_2^2+ B\,d\phi_3^2 \Big] 
- 2 \, \frac{r_0 \, \eta^2 \, z^2}{\sqrt{g}} \, d\tau \Big[A \, d\phi_2 +B\, d\phi_3 \Big]
\end{equation}
with
\begin{equation}
A = \frac{1-\psi_0}{2} \quad \& \quad B = \frac{1+\psi_0}{2}.
\end{equation}
The different functions in \eqref{metric-general-v2} are defined as follows
\begin{equation}
f_1 = 1 - \eta^4 \, z^4 \, , \quad g = 1-r_0^2 \, z^2  \quad \& \quad f_2 = \sqrt{g} - \frac{\eta^4 \, z^4}{\sqrt{g}} 
\end{equation}
where we have performed the analytic continuation $t \rightarrow i \, \tau$ and $\phi \rightarrow - \, i \, \phi$, 
in order to have Euclidean signature in the asymptotically $AdS_5$ boundary.\footnote{The 10-dimensional metric has, of course, Minkowskian signature due to the double analytic continuation just mentioned.  }  When $\omega_2$ and  $\omega_3$ are both positive 
(rotation in the same direction) the value of $\psi_0$ and of the constants $A$ and $B$ become
\begin{equation} \label{ABconstants-positive}
\psi_0 = \frac{-\, \omega_2+\omega_3}{\omega_2+\omega_3} \quad \Rightarrow \quad 
A = \frac{\omega_2}{\omega_2+\omega_3} \quad \& \quad B = \frac{\omega_3}{\omega_2+\omega_3} 
 \,\,\,  {\rm for}  \,\,\,  \omega_2 > 0 \,\,\, \& \,\,\,   \omega_3 > 0 \, . 
\end{equation}
In this case the energy and angular momentum for the rotating string are equal, i.e. $\Delta=J=\omega_2 + \omega_3$.
When one of the two angular momenta is negative (rotation in different directions) the value of  $\psi_0$ 
and of the constants $A$ and $B$ become
\begin{equation}  \label{ABconstants-negative}
\psi_0 =- \,  \frac{\omega_2+\omega_3}{\omega_2-\omega_3} \quad \Rightarrow \quad 
A = \frac{\omega_2}{\omega_2-\omega_3} \quad \& \quad B = - \, \frac{\omega_3}{\omega_2-\omega_3} 
 \,\,\,  {\rm for}  \,\,\,  \omega_2 > 0 \,\,\, \& \,\,\,   \omega_3 < 0 \, . 
\end{equation}
In this case the energy and angular momentum for the rotating string are different \footnote{Notice that the last two equations can be written in a compact expression as \begin{equation} \label{ABconstants-unified}
\psi_0 = \frac{-\, \omega_2+|\omega_3|}{\omega_2+|\omega_3|} \quad \Rightarrow \quad 
A = \frac{\omega_2}{\omega_2+|\omega_3|} \quad \& \quad B = \frac{|\omega_3|}{\omega_2+|\omega_3|} 
 \,\,\,   \, . 
\end{equation} }, i.e. 
\begin{equation} \label{Delta-J-negative}
\Delta = \omega_2 + |\omega_3| = \omega_2 - \omega_3 > 0 \quad \& \quad J = \omega_2 +\omega_3= \omega_2 -|\omega_3|\, . 
\end{equation}

The value of  $\psi_0$ comes from solving the equation of motion for the coordinate $\psi$. There are two solutions for 
$\psi_0$ and in order for $\psi$ to range in the interval $[0,{\pi/2}]$, we assign one of the solutions to the case that both 
angular momenta are positive (i.e. \eqref{ABconstants-positive}) and the other solution to the case that angular momenta 
have different signs (i.e. \eqref{ABconstants-negative}). Notice also that in both cases, if we take the limit 
$\omega_3 \rightarrow 0$, we obtain the solution of \cite{Georgiou:2022ekc}.
Let us stress once more that the aforementioned point-like solutions on the metric \eqref{metric-general-v2} are also solutions of the 10-dimensional metric \eqref{metric-general-v1}.


\section{Calculating correlation functions}
\label{setup}

In this section, we will present our general method that will be used in the holographic calculation of two 
and most importantly higher point correlation functions involving operators that are dual to classical (point-like) string states with large charges.

This method has been already employed in \cite{Georgiou:2022ekc} in order to calculate two-point correlation functions of operators dual to classical string states in the presence of both temperature and R-charge density.  The corresponding dual operators were carrying a single non-zero R-charge which corresponds to point particles rotating along a single isometry of the 5-dimensional deformed internal space. 
The purpose of this section is to extend this approach to accommodate the case of a classical string that rotates along two angles of the internal space, carrying, thus, two angular momenta.  
If one wishes to calculate three-point correlation functions involving the class of solutions presented in \cite{Georgiou:2022ekc}, namely solutions of constant $\psi$,  \footnote{Notice that in the undeformed case it is possible to have non-extremal three-point correlators involving operators in a single $SU(2)$ sector, which involves a single isometry \cite{Buchbinder:2011jr}. In this case one of the corresponding string solutions has a non-constant value for the angle $\psi$. In contradistinction, for the solutions we are using $\psi$ is constant.} this generalisation seems absolutely crucial. The reason is the following: In the case in which only one isometry is turned on,  conservation of the R-charge implies that the dimensions of the operators $\Delta_i,\,\,\,i=1,2,3$ should satisfy the relation $\Delta_1=\Delta_2+\Delta_3$. 
Consequently, the corresponding three-point function will be extremal. But as was already mentioned in \cite{Klose:2011rm}, at the strong coupling regime the extremal three-point correlator factorises into the product of two two-point functions. This happens because the interacting point, at which the three geodesics join, is situated on the boundary of the space-time.\footnote{This was proven to hold in \cite{Klose:2011rm} for the $AdS_5$ space and in \cite{Rodriguez-Gomez:2021mkk} for the case of a thermal CFT. In what follows, we will see that this is also true in the presence of both temperature and chemical potential.
It is not obvious to us if this factorisation persists beyond the strong coupling regime.} 
Consider now the case of strings having two angular momenta. The dual operators will schematically be of the form 
\begin{equation} \label{ops}
{\cal O}_1={\rm sTr}Z_1^{J_1}  Z_2^{J_2}, \,\,\,\quad
{\cal O}_2={\rm sTr}{Z_1^{J_3} {\bar Z}_2^{J_4}}, \,\,\, \quad
{\cal O}_3={\rm sTr}{{\bar Z}_1^{J_5} {\bar Z}_2^{J_6}}
\end{equation} 
of the theory, where ${\rm sTr}$ denotes all possible symmetrisations of the fields inside the trace.  R-charge conservation along the two isometries then implies the relations $J_1+J_3=J_5$ and $J_2=J_4+J_6$. Taking into account that $\Delta_1=J_1+J_2$,  
$\Delta_2=J_3+J_4$ and $\Delta_3=J_5+J_6$ we deduce that $\Delta_2+\Delta_3=\Delta_1+2 J_3>\Delta_1$. Thus, we see that when one of the operators is of the type in \eqref{Delta-J-negative} the three-point correlator will be necessarily non-extremal. The discussion above justifies the need for the generalisation of the string solution compared to the ones used in \cite{Georgiou:2022ekc}.

Consider now a metric that has the following generic form
\begin{equation} \label{metric-general-g2}
ds^2 = g_{tt}\, d\tau^2 + g_{xx}\, d\vec{x}_3^2 +g_{zz} \, dz^2 +
g_{\phi\phi} \, \Big[ A \, d\phi_2^2 + B \, d\phi_3^2\Big]+
2 \, g_{t\phi} \, d\tau \, \, \Big[ A \, d\phi_2 + B \, d\phi_3 \Big] 
\end{equation}
In what follows, and in order to keep the discussion as simple as possible, we will seek point-like solutions that propagate from one boundary point to another with the two boundary points having only temporal distance. The corresponding field theory object is the two-point function of scalar operators with large charges and dimensions. As it usually happens for operators of large dimensions, the path integral is dominated by a classical trajectory and the result for the correlator is given in terms of the length of a geodesic. However there is a main obstacle. In the case where the internal space is not maximally symmetric,  the motion of the particle in the internal space much be correctly taken into account. As a result, the motion of the particle is governed by an effective Lagrangian which comes out once the degrees of freedom of the internal space are integrated. The way to derive this effective Lagrangian has been put forward in \cite{Georgiou:2022ekc}. As mentioned above, we will generalise this method for the case of two angular momenta.

One starts with the Polyakov action $S_P$ for the string and then performs a Legendre transformation with respect to the two isometries $\phi_2$ and $\phi_3$, in order to pass from the variables 
$(\phi_i,\dot\phi_i)$ to  $(\phi_i,{\cal\pi_\phi}_i)$. Here ${\cal\pi_\phi}_i$ is the momentum conjugate to the isometry $\phi_i, \,i=2,3$ and is a constant of the motion. 
In particular one has 
\begin{equation}  \label{NG}
 S=S_P-\int d^2 \sigma \,{\pi_\phi}_i \dot\phi_i=
 \int d^2 \sigma {\cal L}-\int d^2 \sigma \,{\pi_\phi}_i \dot\phi_i\,.
\end{equation}
The last term in the action acts as a wave functional projector imposing the fact that initial and final states have well-defined fixed angular momenta.
One may  solve the equations of motion for the isometries $\tau$, $\phi_2$ and $\phi_3$ to get
\begin{equation}   \label{eom1}
\pi_{\tau}\, = \, \frac{\partial \cal L}{\partial \dot \tau} \, = \, \mu \,\Delta  
\quad \& \quad 
\pi_{\phi_{2,3}} \, = \, \frac{\partial \cal L}{\partial \dot \phi_{2,3}} \, = \, \omega_{2,3}
\end{equation}
where $\mu$ and $\omega_{2,3}$ are constants.
Let us mention that the derivative of $\tau$, $\phi_2$ and $\phi_3$ is with respect to the world-sheet time. 
The above set of three equations can now be solved for 
$\dot \tau =\dot \tau(z,\Delta, \omega_2,\omega_3)$ and $\dot \phi_{2,3}=\dot \phi_{2,3}(z,\Delta, \omega_2,\omega_3)$.

Subsequently, after having substituted in the non-trivial Virasoro constraint the aforementioned expressions for $\dot \tau$ and $\dot \phi_{2,3}$, one solves this constraint, namely $g_{\mu\nu} \partial_\tau X^\mu \partial_\tau X^\nu=0$ for $\dot z=\dot z(z,\Delta, \omega_2,\omega_3)$ and substitute this result in the action \eqref{NG}. In this way one ends up with an action that depends only on the conserved charges, the dimension of the operator $\Delta$ and the holographic coordinate $z$. 

The virtue of \eqref{NG} is that we have integrated out the coordinates of the 5-dimensional internal space. To summarise,
the effective action \eqref{NG} describes the motion of a point-like string in the 5-dimensional space-time 
$(\tau,z,\vec x)$ with the motion of the particle in the sphere taken properly into account. At this point we choose the gauge in which the world-sheet time is identified with $z$, which implies that from now on the dot will denote derivative with respect to $z$.

The equation that determines the temporal boundary distance when both $\omega_2$ and  $\omega_3$ are positive is given by 
\begin{equation}\label{td-positive}
\dot \tau \, = \, i \, \sqrt{\big. g_{zz} \, g_{xx}}\frac{g_{t\phi} -\mu \,g_{\phi\phi}}{D_p}
\end{equation}
where $\mu$ is a constant and $D_p$ is defined as follows
\begin{equation} \label{D-def-positive}
D_p^2 \, = \, \left(g_{t\phi}^2 - g_{\phi \phi} g_{tt} \right)\, g_{xx} 
\Big[- g_{tt}- \mu^2 \, g_{\phi \phi} + 2 \, \mu \, g_{t\phi} \Big]\, .
\end{equation}
Similar expressions can be derived for $\dot\phi_2$ and $\dot\phi_3$ which we refrain from presenting here.
The action after substituting the expressions for $\dot \tau$, $\dot \phi_2$ and $\dot \phi_3$ becomes
\begin{equation} \label{NGfin-positive}
S \, = \, i \, \Delta \int dz \sqrt{\big.g_{zz} \, g_{xx}}  \, \frac{g_{tt}-\mu \, g_{t \phi}}{D_p} \, .
\end{equation}
When one of the two angular momenta is negative (i.e. $\omega_3 <0$) the equation determining the temporal boundary distance is given by a different expression, namely 
\begin{equation}\label{td-negative}
\dot \tau \, = \, i \, \sqrt{\big. g_{zz} \, g_{xx}}\frac{\frac{J}{\Delta} \, g_{t\phi} -\mu \,g_{\phi\phi}}{D_n}
\end{equation}
where $D_n$ is defined as follows
\begin{equation} \label{D-def-negative}
D_n^2 \, = \, \left(g_{t\phi}^2 - g_{\phi \phi} g_{tt} \right)\, g_{xx} 
\Bigg[- g_{tt} - \mu^2 \,  g_{\phi \phi} +  2 \, \frac{J}{\Delta} \,\mu \, g_{t\phi} + \left(1 - \frac{J^2}{\Delta^2} \right) 
\frac{g_{t\phi}^2}{g_{\phi \phi}} \Bigg]\, .
\end{equation}
Similar expressions can be derived for $\dot\phi_2$ and $\dot\phi_3$ which we refrain from presenting here.
The constants  $\omega_2$ and  $\omega_3$ are related to the conformal dimension and angular momenta as shown in
\eqref{Delta-J-negative}.  Substituting in the expression for the action we obtain
\begin{equation} \label{NGfin-negative}
S \, = \, i \, \Delta \int dz \sqrt{\big.g_{zz} \, g_{xx}}  \, \frac{g_{tt}- \frac{J}{\Delta} \, \mu \, g_{t \phi} - 
 \left(1 - \frac{J^2}{\Delta^2} \right) 
\frac{g_{t\phi}^2}{g_{\phi \phi}} }{D_n} \, .
\end{equation}
An important comment is in order. Notice that in contradistinction to \eqref{NGfin-positive} the action \eqref{NGfin-negative} depends not only on the conformal dimension of the operator but also on the difference of the two angular momenta. Compared to the calculation of correlators in other holographic set-ups  this is a novel feature that is due to the fact the background we are working with distinguishes the direction of rotation. It is not invariant under $\phi_{2,3}\rightarrow -\phi_{2,3}$ as it happens to all backgrounds considered in the literature so far.

Building on the approach that we have just reviewed, one can move forward to the holographic calculation of 
the following three-point correlation function
\begin{equation}
G_3 \, = \, \langle{\cal O}_{\Delta_1}(\tau_1) {\cal O}_{\Delta_2}(\tau_2){\cal O}_{\Delta_3}(\tau_3)\rangle \,  .
\end{equation}
For large values of the conformal dimensions $\Delta_1$, $\Delta_2$ and $\Delta_3$ the three-point correlator 
will be dominated by three geodesic arcs that join at the bulk point $(\tau, z)=(\tau_I, z_I)$. 
This point will be determined by demanding that the total action $S= S_1 + S_2 + S_3$ is an extremum, that is
\begin{eqnarray} \label{ti-zi}
&&\frac{dS}{d\tau_I} \, = \, 0 \quad \Rightarrow  \quad \sum_{i=1}^3 \pi_t^{(i)} \, = \, 0  \quad \Leftrightarrow \quad 
\sum_{i=1}^3\mu_i \, \Delta_i \, = \, 0
\nonumber \\
&&\frac{dS}{dz_I} \, = \, 0 \quad \Rightarrow  \quad  \sum_{i=1}^3 \epsilon_i \,  \Delta_i  \, D(\mu_i, z_I) \, = \, 0
\end{eqnarray}
where $\epsilon_i = 1$ if the corresponding geodesic has not reached the point where $\frac{dz}{d\tau}=0$
(an {\it ingoing} geodesic in the terminology of \cite{Rodriguez-Gomez:2021mkk}) 
while $\epsilon_i = -1$ if the corresponding geodesic has passed this point and is in its way to return to the boundary
(a {\it returning} geodesic in the terminology of \cite{Rodriguez-Gomez:2021mkk}).
The contribution to the string action for an ingoing geodesic is
\begin{eqnarray} \label{S-ingoing}
S_{\rm in} \, = \, S_{\tau}(z_I)
\quad &{\rm with}& \quad 
S_{\tau}(z) \, = \, i \, \Delta \, \int_{\epsilon}^{z} dz \sqrt{\big.g_{zz} \, g_{xx}}  \, \frac{g_{tt}-\mu \, g_{t \phi}}{D_p} 
\\ 
\quad  &{\rm or}& \quad  S_{\tau}(z)  \, = \, i \, \Delta  \int_{\epsilon}^{z} 
dz \sqrt{\big.g_{zz} \, g_{xx}}  \, \frac{g_{tt}- \frac{J}{\Delta} \, \mu \, g_{t \phi} - 
 \left(1 - \frac{J^2}{\Delta^2} \right) 
\frac{g_{t\phi}^2}{g_{\phi \phi}} }{D_n}
\nonumber 
\end{eqnarray}
while for a returning one is
\begin{equation}  \label{S-returning}
S_{\rm ret} \, = \, 2 \,S_{\rm max} \, - \, S_{\rm ing}
\quad {\rm with} \quad 
S_{\rm max} \, = \, S_{\tau}(z_{\rm max})
\end{equation}
where $z_{\rm max}$ is the maximum point that the geodesic penetrates into the bulk. At this point, for temporal (spatial) 
boundary distance the derivative $dz/d\tau$ ($dz/dx$) vanishes.
The first equation in \eqref{ti-zi} is just the conservation of momentum along the $\tau$-isometry. 
The second equation can be derived by taking into account that $S$ depends on $z_I$ both explicitly and implicitly 
through the dependence of $\mu_i$ on $z_I$. 
The equation that describes an incoming geodesic connecting the point $(\tau_i, 0)$ on the boundary 
with the meeting point $(\tau_{I},z_{I})$ in the bulk is
\begin{eqnarray} \label{incoming-geodesic}
\tau_{I}^{\rm ingoing} \, = \, \tau_i \,  + \, I_{\tau}(z_I)
\quad  &{\rm with}& \quad  I_{\tau}(z)  \, = \, i \,  \int_0^{z}  dz  \sqrt{\big. g_{zz} \, g_{xx}}\frac{g_{t\phi} -\mu \,g_{\phi\phi}}{D_p}
\\ 
\quad  &{\rm or}& \quad  I_{\tau}(z)  \, = \, 
i \, \int_0^{z}  dz \, \sqrt{\big. g_{zz} \, g_{xx}}\frac{\frac{J}{\Delta} \, g_{t\phi} -\mu \,g_{\phi\phi}}{D_n}
\nonumber 
\end{eqnarray}
and the equation that describes a returning geodesic connecting the same points is 
\begin{equation} \label{returning-geodesic}
\tau_{I}^{\rm returning}\, =  \, 2 \, \tau_{\rm max}\, - \, \tau_{I}^{\rm ingoing} 
\quad {\rm with} \quad 
\tau_{\rm max} \, = \, \tau_i  + I_{\tau}(z_{\rm max}) \, . 
\end{equation}

To determine the three-point function, an algebraic system of five equations and five unknowns has to solved. 
The five unknowns are:  the three conserved momenta $\mu_1$, $\mu_2$ and $\mu_3$ and the coordinates, 
$\tau_I$ and $z_I$, of the meeting point. The five equations are: the two equations from extremising the action in 
\eqref{ti-zi} and the three geodesics from the three operators on the boundary until  the meeting point  $(\tau_I, z_I)$ 
in the bulk. These geodesics can be either 2 incoming and 1 returning or 1 incoming and 2 returning 
and are described by equations \eqref{incoming-geodesic} and \eqref{returning-geodesic}.
The final expression for the holographic three point correlator will then be given by the on-shell action, namely
\begin{equation} \label{3-point-gen}
G_3 \, = \, \langle{\cal O}_{\Delta_1}(\tau_1) {\cal O}_{\Delta_2}(\tau_2){\cal O}_{\Delta_3}(\tau_3)\rangle ={\cal G}_{\Delta_1\Delta_2\Delta_3} \exp \left[{-\sum_{i=1}^3 S_i(\tau_1,\tau_2,\tau_3)} \right] \, ,
\end{equation}
where ${\cal G}_{\Delta_1\Delta_2\Delta_3}$ is the supergravity coupling of the three states with large charges.

\section{Two-point correlation functions}
\label{2point-functions}

In this section we will take one step backwards and focus attention on the calculation of two-point correlation functions. 
Since in the previous section we have introduced a new solution the 
first natural step is to study the behaviour of the two-point correlator. Here we refer to the case with negative $\omega_3$, 
because for the case with positive $\omega_3$, see \eqref{td-positive} to \eqref{NGfin-positive}, the analysis is identical to the case with one angular momentum that was explored in \cite{Georgiou:2022ekc}. This is so because in this case the dependence of the action on the internal space enters explicitely only through the dimension of the operators $\Delta$. As we have seen above, this is no longer true for the case with $\omega_3<0$.

In order to be self contained we also include the case with
$\omega_3>0$. This is presented in subsection \ref{2point-positive}, and it contains results from subsection 6.2.1 of 
\cite{Georgiou:2022ekc}. The case with $\omega_3<0$ is presented in subsection \ref{2point-negative}. Notice that we consider
correlators with only temporal boundary distance (the calculation can easily be extended to incorporate spatial boundary distance),
while the two-point function depends on both temperature and chemical potential. It should be stressed that only in this case 
the two solutions we presented in the previous section differ one from the other, since only then the non-diagonal term in the metric is different
from zero. In any other case, i.e. either with  temperature only or with  R-charge only, the two solutions become identical, 
and the respective two-point functions have been computed in \cite{Rodriguez-Gomez:2021pfh} and \cite{Georgiou:2022ekc}, respectively.


\subsection{Rotation in the same direction}
\label{2point-positive}

Substituting the following values for the metric components 
\begin{equation} \label{metric_components}
g_{tt} = \frac{f_2}{z^2} \, , \quad
g_{xx} = \frac{\sqrt{g}}{z^2} \, , \quad
g_{zz} = \frac{1}{z^2}\frac{1}{\sqrt{g} \, f_1}  \, , \quad
g_{\phi \phi} = - \, \frac{1}{\sqrt{g}} \quad \& \quad 
g_{t \phi} = - \, \frac{r_0 \, \eta^2 \, z^2}{\sqrt{g}}
\end{equation}
in \eqref{td-positive} for the temporal boundary distance and in \eqref{NGfin-positive} for the action we arrive to the 
following expressions
\begin{equation} \label{2point-tdot-positive-v1}
\dot \tau \, = \, \frac{\mu \, z - r_0 \, \eta^2 \, z^3}{\left(1 - \eta^4 \, z^4\right)\sqrt{\big. \left(1 - r_0^2 \, z^2\right)h(z) }} 
\quad {\rm and} \quad  
\frac{S}{\Delta} \, = \,  \int \frac{dz}{z} \,\frac{1 - r_0^2 \, z^2 - \eta^4 \, z^4 +r_0 \, \eta^2 \, \mu \, z^4}{\left(1 - \eta^4 \, z^4\right)\sqrt{\big. \left(1 - r_0^2 \, z^2\right)h(z) }} 
\end{equation}
with the function $h(z)$ having the following expression
\begin{equation}
h(z) = 1 -\eta^4 \, z^4 - \left(\mu^2 + r_0^2 \right) z^2 +2 \, r_0 \, \eta^2 \, \mu \, z^4 \, . 
\end{equation}
The turning point for the string inside the bulk is given by the point that the function $h(z)$ vanishes, namely
\begin{equation}
h(z)= 0  \quad \Rightarrow \quad 
z_{\rm max}^2 = - \, \frac{\mu ^2+r_0^2 -
\sqrt{\big.4 \, \eta ^4+\mu ^4-8 \, \eta ^2 \, \mu  \, r_0+2 \, \mu ^2 \, r_0^2 + r_0^4}}{2 \, \eta ^4-4 \, \eta ^2 \, \mu  \, r_0} \, . 
\end{equation}
Equation \eqref{2point-tdot-positive-v1} should be solved with the boundary conditions $\tau(z=0)=-\tau$ and $\tau(z=z_{\rm max})=0$ which implies that the operator insertions are at $(-\tau,0)$ and $(\tau,0)$, respectively.
Solving perturbatively the integrals in \eqref{2point-tdot-positive-v1} in the limit of nearly coincident operators, 
i.e. 
$\tau \ll \eta^{-1}$
or equivalently $\mu \rightarrow \infty$, we obtain the following result 
for the temporal distance\footnote{In what follows we have re-scaled the temporal distance $\tau$ as follows: $\tau \rightarrow \frac{\tau}{2}$. That means that one operator is now at $(-\frac{\tau}{2}, \vec{0})$ and the second operator at $(\frac{\tau}{2}, \vec{0})$.}
\begin{eqnarray}  \label{2point-t-positive_v1}
&& \frac{\mu \, \tau}{2} = 1-\frac{2 \, r_0^2}{3 \, \mu ^2}+\frac{2 \, \eta ^2 \, r_0}{\mu ^3}+
\frac{4 \left(2 \, r_0^4-3 \, \eta ^4\right)}{15 \, \mu ^4}-
\frac{16 \left(\eta ^2 \, r_0^3\right)}{3 \, \mu ^5}-
\frac{16 \left(r_0^6-25 \, \eta ^4 \, r_0^2\right)}{35 \, \mu^6}
\nonumber \\[5pt]
&& 
\qquad \qquad
+\frac{48 \, \eta ^2 \, r_0^5-40 \,  \eta ^6 \, r_0}{5\, \mu ^7}+
\frac{16 \left(35 \, \eta ^8-924 \, \eta ^4 \, r_0^4+8 \, r_0^8\right)}{315 \, \mu ^8}
\end{eqnarray}
while the on-shell action becomes
\begin{equation} \label{2point-action-positive_v1}
- \frac{2 \, S_{os}}{\Delta} = 
2 \Bigg[\log \left(\frac{\epsilon \, \mu}{2} \right)+
\frac{r_0^2}{\mu ^2} - 
\frac{8 \, \eta ^2 \, r_0}{3 \, \mu ^3} +
\frac{3 \, \eta ^4-2 \, r_0^4}{3 \, \mu ^4}+ 
\frac{32 \, \eta ^2 \, r_0^3}{5 \, \mu ^5} - 
\frac{8 \left(25 \, \eta ^4 \, r_0^2- r_0^6\right)}{15 \, \mu ^6} \Bigg] \, . 
\end{equation}
Combining \eqref{2point-t-positive_v1} and \eqref{2point-action-positive_v1}  we calculate the 
two-point function as a series expansion for small values of $\tau$, or equivalently  for small values of $r_0$ and $\eta$
\begin{eqnarray}  \label{2point-action-positive_v2}
\frac{1}{\Delta}\log \langle{\cal O}_{\Delta}(0) {\cal O}_{\Delta}(\tau)\rangle &=& - \, 2 \, \log \tau  + 
\frac{1}{6} \, r_0^2 \, \tau ^2 -
\frac{1}{6} \, r_0 \, \eta^2 \, \tau ^3  +
\frac{1}{90} \, r_0^4 \, \tau ^4 + 
\frac{1}{40} \, \eta^4 \, \tau^4 
\nonumber \\[5pt]
&& 
-  \, \frac{1}{60} \, r_0^3 \, \eta^2 \, \tau^5 + 
\frac{47}{22680} \, r_0^6 \, \tau^6 + 
\frac{11}{560} \, r_0^2 \, \eta^4 \, \tau^6 \, . 
\end{eqnarray}
Notice that the leading term in the last equation is the result for the two-point function in a conformal field theory.


\subsection{Rotation in opposite directions}
\label{2point-negative}

Now we move to the case where the rotation in the internal space for $\phi_2$ and $\phi_3$ is in opposite directions. Without 
loss of generality we have chosen $\omega_2>0$ and $\omega_3<0$, so substituting the values of the metric 
components \eqref{metric_components} in equations \eqref{td-negative} and \eqref{NGfin-negative}, 
we arrive to the following expression governing the temporal boundary distance
\begin{equation}  \label{2point-tdot-negative-v1}
\dot \tau \, = \, \frac{\mu \, z - \frac{J}{\Delta} \, r_0 \, \eta^2 \, z^3}{\left(1 - \eta^4 \, z^4\right)
\sqrt{\big. \left(1 - r_0^2 \, z^2\right)h(z) }}
\end{equation}
and the on-shell action
\begin{equation} \label{2point-action-negative-v1}
S \, = \, \Delta \int \frac{dz}{z} \,\frac{1 - r_0^2 \, z^2 - \eta^4 \, z^4 +\frac{J}{\Delta} \, r_0 \, \eta^2 \, \mu \, z^4
+ \left(1- \frac{J^2}{\Delta^2}\right) r_0^2 \, \eta^4 \, z^6}{\left(1 - \eta^4 \, z^4\right)\sqrt{\big. \left(1 - r_0^2 \, z^2\right)h(z) }}\, .  
\end{equation}
The function $h(z)$ has the following expression
\begin{equation}
 h(z) = 1 -\eta^4 \, z^4 - \left(\mu^2 + r_0^2 \right) z^2 +\frac{2\, J}{\Delta}\, r_0 \, \eta^2 \, \mu \, z^4 
+ \left(1- \frac{J^2}{\Delta^2}\right) r_0^2 \, \eta^4 \, z^6
\end{equation}
while the relation between $\omega_2$ and $\omega_3$ with the angular momentum $J$ and the conformal dimension $\Delta$
is given in \eqref{Delta-J-negative}. To obtain the turning point for the string world-sheet in the bulk 
we solve perturbatively the equation $h(z)= 0$ for large values of $\mu$, namely
\begin{eqnarray}
h(z)= 0  \quad \Rightarrow \quad 
\mu\, z_{\rm max}  &=& 1-\frac{r_0^2}{2 \, \mu ^2}+ \frac{J}{\Delta}\, \frac{\eta ^2 \, r_0}{\mu ^3}+
\frac{3 \, r_0^4-4 \, \eta ^4}{8 \, \mu ^4} - \frac{J}{\Delta} \, \frac{5 \, \eta ^2 \, r_0^3}{2 \, \mu ^5}
\\[5pt]
&&+ \, \frac{r_0^2}{16 \, \mu ^6} \left[4 \, \eta ^4 \left(\frac{12 \, J^2}{\Delta ^2}+7 \right)-5 \, r_0^4\right] +\frac{J}{\Delta} \, 
\frac{7 \left(5 \, \eta ^2 \, r_0^5-4 \, \eta ^6 \, r_0\right)}{8 \, \mu^7} \, . 
\nonumber
\end{eqnarray}
As in the previous subsection, we solve perturbatively the integrals in \eqref{2point-tdot-negative-v1} and 
\eqref{2point-action-negative-v1} in the limit of $\mu \rightarrow \infty$. Rescaling the temporal distance as 
$\tau \rightarrow \frac{\tau}{2}$, we have 
\begin{eqnarray} \label{2point-t-negative_v1}
\frac{\mu \, \tau}{2} &=& 1-\frac{2 \, r_0^2}{3 \, \mu ^2}+\frac{J}{\Delta} \,  \frac{2 \, \eta ^2 \, r_0}{\mu ^3}
+\frac{4 \left(2 \, r_0^4-3 \, \eta ^4\right)}{15 \, \mu ^4} - \frac{J}{\Delta} \,  \frac{16 \, \eta ^2 \, r_0^3}{3\,  \mu^5}
 \nonumber \\[5pt]
&& 
+ \, \frac{8 \, r_0^2}{35 \, \mu ^6} \left[5 \, \eta ^4 \left(\frac{7 \, J^2}{\Delta ^2}+3 \right)-2 \, r_0^4\right]
+\frac{J}{\Delta} \,  \frac{8 \, \eta ^2 \, r_0 \left(6 \, r_0^4-5 \, \eta ^4\right)}{5 \, \mu ^7}
\end{eqnarray}
while for the action we get
\begin{eqnarray}  \label{2point-action-negative_v1}
- \frac{2 \, S_{os}}{\Delta} &=& 
2 \Bigg[\log \left(\frac{\epsilon \, \mu}{2} \right)+
\frac{r_0^2}{\mu ^2} - 
\frac{J}{\Delta} \,  \frac{8 \, \eta ^2 \, r_0}{3 \,  \mu ^3}+ 
\frac{\eta^4-\frac{2 \, r_0^4}{3}}{\mu ^4}+
\frac{J}{\Delta} \,  \frac{32 \, \eta ^2 \, r_0^3}{5\,  \mu^5}
 \nonumber \\[5pt]
&& + \, 
\frac{r_0^2}{3\, \mu^6} \, \left[\frac{8 \, r_0^4}{5}- 4\,  \eta ^4 
\left(\frac{7 \, J^2}{\Delta^2}+3\right) \right]+
\frac{J}{\Delta} \,  \frac{64 \, \eta ^2 \, r_0 \left(5 \, \eta ^4-6 \, r_0^4\right)}{35 \, \mu ^7}\Bigg] \, . 
\end{eqnarray}
The final step is to combine \eqref{2point-t-negative_v1} and \eqref{2point-action-negative_v1} to calculate the 
two-point function as a series expansion  for small values of $r_0$ and $\eta$. The result reads
\begin{eqnarray} \label{2point-action-negative_v2}
\frac{1}{\Delta}\log \langle{\cal O}_{\Delta}(0) {\cal O}_{\Delta}(\tau)\rangle &=& - \, 2 \, \log \tau  + 
\frac{1}{6} \, r_0^2 \, \tau ^2 -
\frac{1}{6} \, \frac{J}{\Delta} \, r_0 \, \eta^2 \, \tau ^3  +
\frac{1}{90} \, r_0^4 \, \tau ^4 + 
\frac{1}{40} \, \eta^4 \, \tau^4 
\\[5pt]
&& 
-  \, \frac{1}{60} \, \frac{J}{\Delta} \,r_0^3 \, \eta^2 \, \tau^5 + 
\frac{47}{22680} \, r_0^6 \, \tau^6 - 
\frac{1}{840} \left[1-\frac{35 \, J^2}{2 \, \Delta ^2}\right] \, r_0^2 \, \eta^4 \, \tau^6 \, . 
\nonumber 
\end{eqnarray}
At this point, it is important to notice that the novel holographic two-point function  \eqref{2point-action-negative_v2} depends not only on the dimension of the operators but also on the difference of their angular momenta. This is the novel feature of our solutions.
Notice also that the expressions for the two-point functions in equations \eqref{2point-action-positive_v2} 
and \eqref{2point-action-negative_v2} coincide when $J=\Delta$. 
Finally, notice that using \eqref{thermo-quantities} it is straightforward to express the result for the two-point function in terms of the field theory quantities $T$ and $\Omega$.


\section{Three-point correlation functions}
\label{3point-functions}

In this section we will present the holographic calculation for the three-point correlators. 
We follow the methodological approach of \cite{Janik:2010gc} 
with the starting point being the effective action that one obtains after the 
Legendre transformation of the Polyakov action and the subsequent use of the Virasoro constraint \cite{Georgiou:2022ekc}.
Those steps have been thoroughly explained in section \ref{setup}.

In the analysis to follow, there are three cases that we will cover. In the first two cases, presented in subsections \ref{3point-Rcharge} and  \ref{3point-Temperarure}, we calculate 
the three-point correlation function when the background contains either only R-charge density or only temperature, 
but not both. Notice that in neither of these two cases there is distinction between a string solution rotating 
in the internal space with the two angular momenta along the same direction or along opposite directions. 
The difference in the behavior between the solutions with values of the constants $A$ and $B$ given in 
\eqref{ABconstants-positive} or given in \eqref{ABconstants-negative}, is based on the presence of the non-diagonal term
in the metric which ceases to exist when either the temperature or the chemical potential is zero. If one of the parameters vanish, the two solutions coincide. This can be also seen by comparing the expressions 
\eqref{2point-tdot-positive-v1} with the equations \eqref{2point-tdot-negative-v1} and \eqref{2point-action-negative-v1}, i.e. 
for $r_0=0$ or $\eta =0$ they coincide. 

Our most general results are presented in subsection \ref{3point-RT} where we calculate the three point correlator when the background is in full generality and contains both 
finite chemical potential and finite temperature. 

\subsection{R-charge density}
\label{3point-Rcharge}

In this case, the expressions for  $\dot \tau$ and the action are given by
\begin{equation}  \label{3point-R-tdot-action}
\dot \tau \, = \, \frac{\mu \, z}{\sqrt{\big.h(z)\left(1- r_0^2 \, z^2\right)}}
\quad \& \quad 
\frac{S}{\Delta} = \int \frac{dz}{z} \, \sqrt{\frac{1- r_0^2 \, z^2}{ h(z) }}
\quad {\rm with} \quad h(z) =   1- \left(\mu^2+ r_0^2\right) z^2
\end{equation}
and can be easily obtained either from \eqref{2point-tdot-positive-v1} or from 
\eqref{2point-tdot-negative-v1} and \eqref{2point-action-negative-v1}, after setting to zero the value of the parameter $\eta$. 
Those integrals can be calculated analytically and will  provide the expressions for $I_{\tau}(z)$ and 
$S_{\tau}(z)$, respectively. The integrations are from the boundary point $z=0$ at which the operator is inserted up to an arbitrary point $z$ in the bulk. More precisely, we obtain
\begin{equation} \label{3point-R-t_v1}
I_{\tau}(z) = \frac{\mu}{2 \, r_0 \sqrt{\big.r_0^2 + \mu^2}} \, \log \left[\frac{\sqrt{\big.r_0^2 + \mu^2} +r_0}{\sqrt{\big.r_0^2 + \mu^2}-r_0} \, \frac{\sqrt{\big. \left(1 -r_0^2 \, z^2\right) \left(r_0^2 + \mu^2\right)} - r_0 \, \sqrt{h(z)} }{\sqrt{\big. \left(1 -r_0^2 \, z^2\right) \left(r_0^2 + \mu^2\right)} + r_0 \, \sqrt{h(z)} }\right]
\end{equation}
and 
\begin{eqnarray}  \label{3point-R-S_v1}
&&\frac{S_{\tau}(z)}{\Delta} = \frac{\mu}{\sqrt{\big.r_0^2 + \mu^2}} \, \log \left[\frac{\sqrt{\big.r_0^2 + \mu^2} -r_0}{\sqrt{\big. \left(1 -r_0^2 \, z^2\right) \left(r_0^2 + \mu^2\right)} - r_0 \, \sqrt{h(z)} }\right]
\\
&& 
+ \frac{1}{2} \log \left[\frac{4}{\epsilon^2 \, \mu^2}\frac{\sqrt{\big.r_0^2 + \mu^2} +r_0}{\sqrt{\big.r_0^2 + \mu^2}-r_0} \, \,
\frac{1-r_0^2 \, z^2 \sqrt{\big.r_0^2 + \mu^2} - \sqrt{\big. h(z) \left(1 - r_0^2 \, z^2\right)} }{1+r_0^2 \, z^2 \sqrt{\big.r_0^2 
+ \mu^2} +\sqrt{\big. h(z) \left(1 - r_0^2 \, z^2\right)}} \right]
\, . \nonumber 
\end{eqnarray}
The value of $z_{\rm max}$ is determined from the point that $dz/d\tau$ vanishes
\begin{equation} \label{3point-R-meeting-point}
\frac{d z}{d \tau }= \frac{1}{\dot \tau} = 0 
 \quad \Rightarrow \quad 
h(z_{\rm \max})= 0  \quad \Rightarrow \quad 
z_{\rm max}^2 = \frac{1}{\mu^2 + r_0^2}\, .
\end{equation}
while the expressions for $ I_{\tau}(z_{\rm max})$ and $S_{\tau}(z_{\rm max})$, that are useful in order to 
construct the equations of the geodesics, read
\begin{equation}
I_{\tau}(z_{\rm max}) = \frac{\mu}{r_0 \sqrt{\big. \mu ^2+r_0^2}} \, \log \left[\frac{\mu }{\sqrt{\big. \mu ^2+r_0^2}-r_0}\right]
\quad \& \quad 
\frac{S_{\tau}(z_{\rm max})}{\Delta} = - \log \frac{\mu}{2}- \frac{r_0^2}{\mu} \, I_{\tau}(z_{\rm max})  \, . 
\end{equation}

\begin{figure}[ht] 
   \centering
   \includegraphics[width=7.5cm]{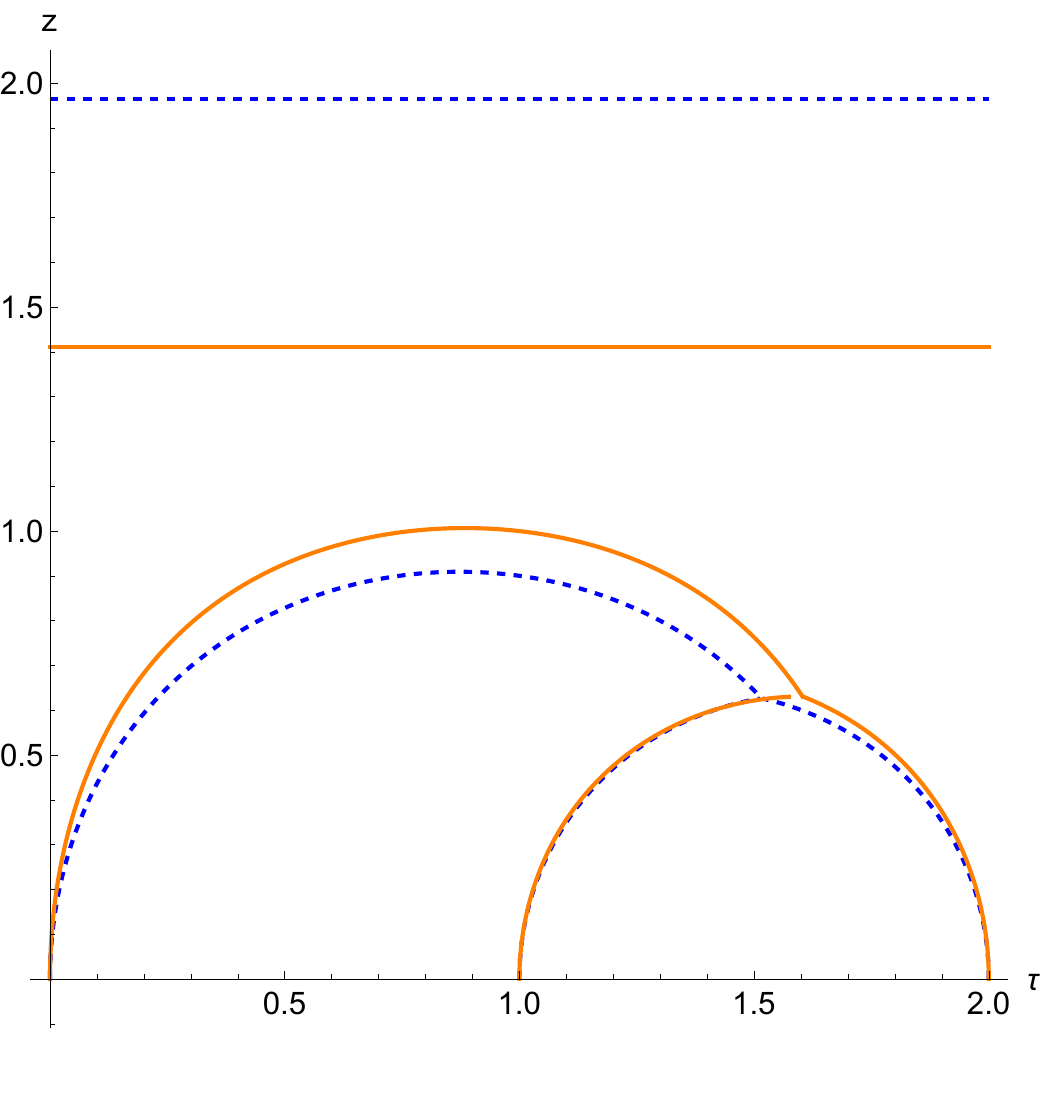}
   \includegraphics[width=7.5cm]{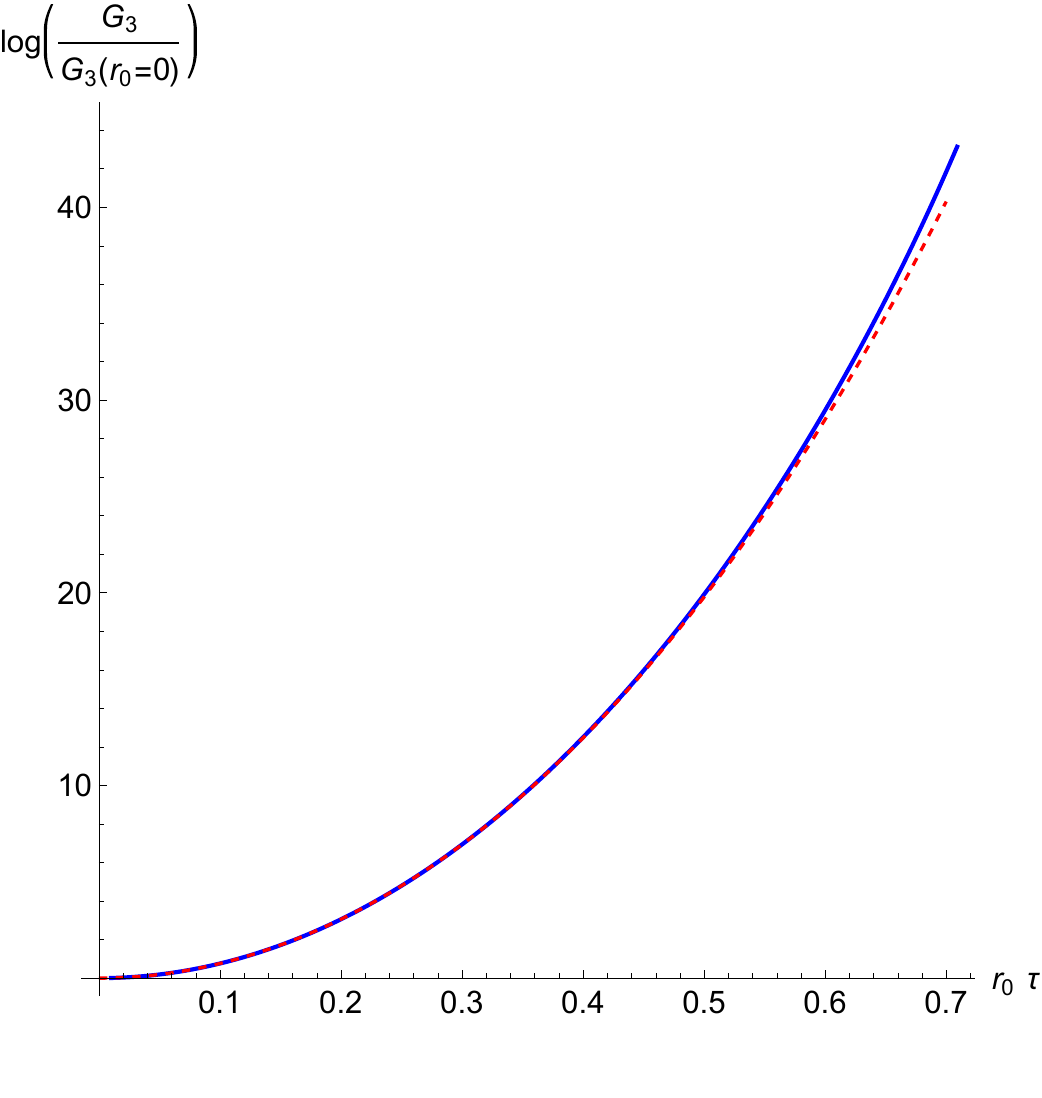}
    \caption{{\bf 2} in going and {\bf 1} returning geodesics for operators with conformal dimensions $\Delta_1= 95$, 
    $\Delta_2= 135$ and $\Delta_3= 200$ at the positions $\tau_1=0$, $\tau_2=1$ and $\tau_3=2$ of the boundary: 
    On the left part we plot the world-sheet of the three-point function. 
    The orange curve corresponds to $r_0=0.709$ and the blue (dashed) curve to $r_0=0.509$.
    The two straight lines determine the position of the singularity for each one of the two values of the parameter $r_0$, 
    i.e. $z_{\rm singular}=r_0^{-1}$. 
    On the right part, we plot the logarithm of the three point function for arbitrary value of the parameter $r_0$ 
    over the three point function for $r_0=0$ as a function of the dimensionless quantity $r_0 \tau $. In the current case we
     have chosen a set-up for which $\tau_3=2 \, \tau_2$ and $\tau_2=1$. As a result, the parameter that changes in the 
     horizontal axis is the value of $r_0$. The blue curve is the result of the numerical 
    computation and for the red (dashed) curve we have used \eqref{3point-R-action-perturbative} 
    that contains the first and the second order corrections to the parameter $r_0$.}
   \label{3point-R-21}
\end{figure}

\begin{figure}[ht] 
   \centering
   \includegraphics[width=7.5cm]{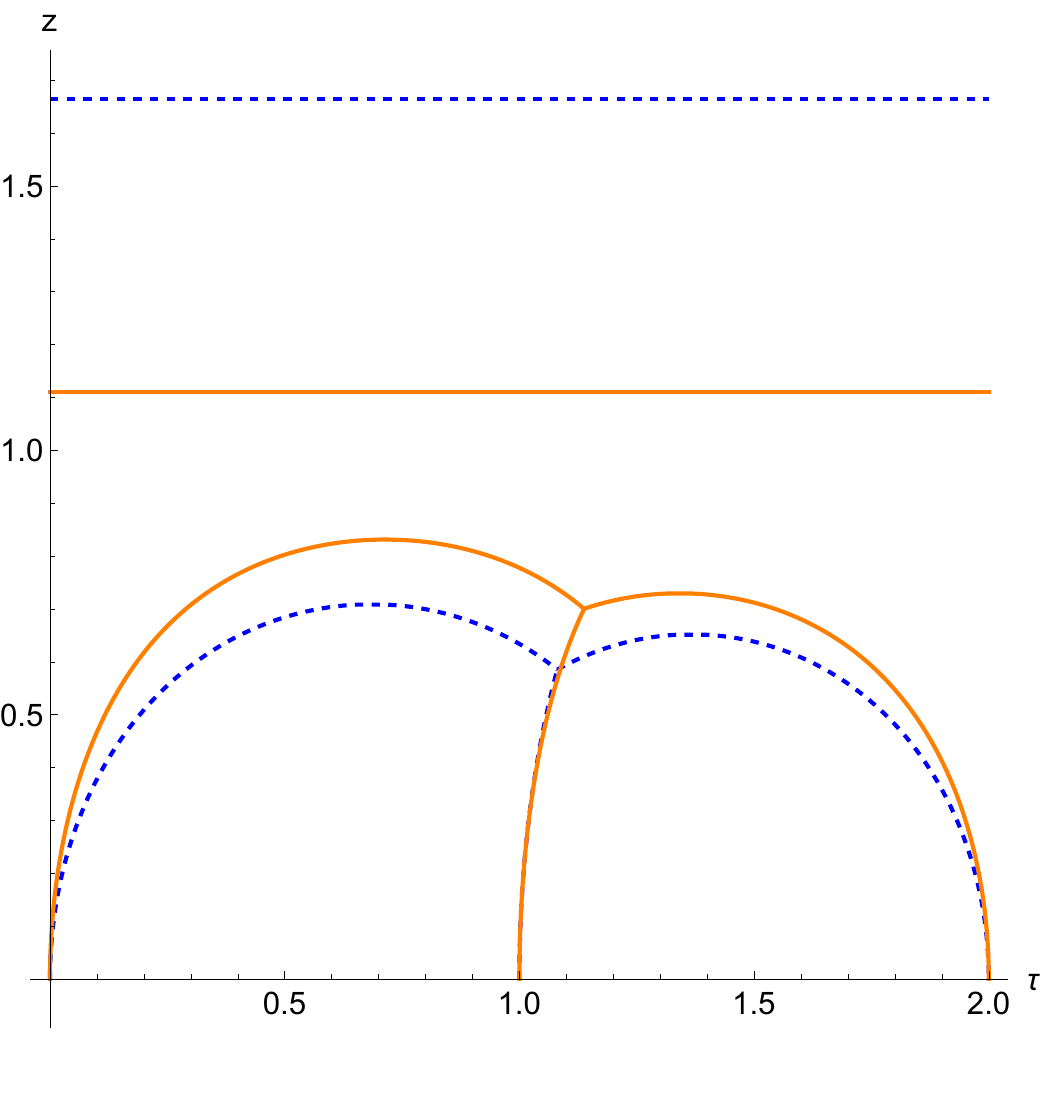}
   \includegraphics[width=7.5cm]{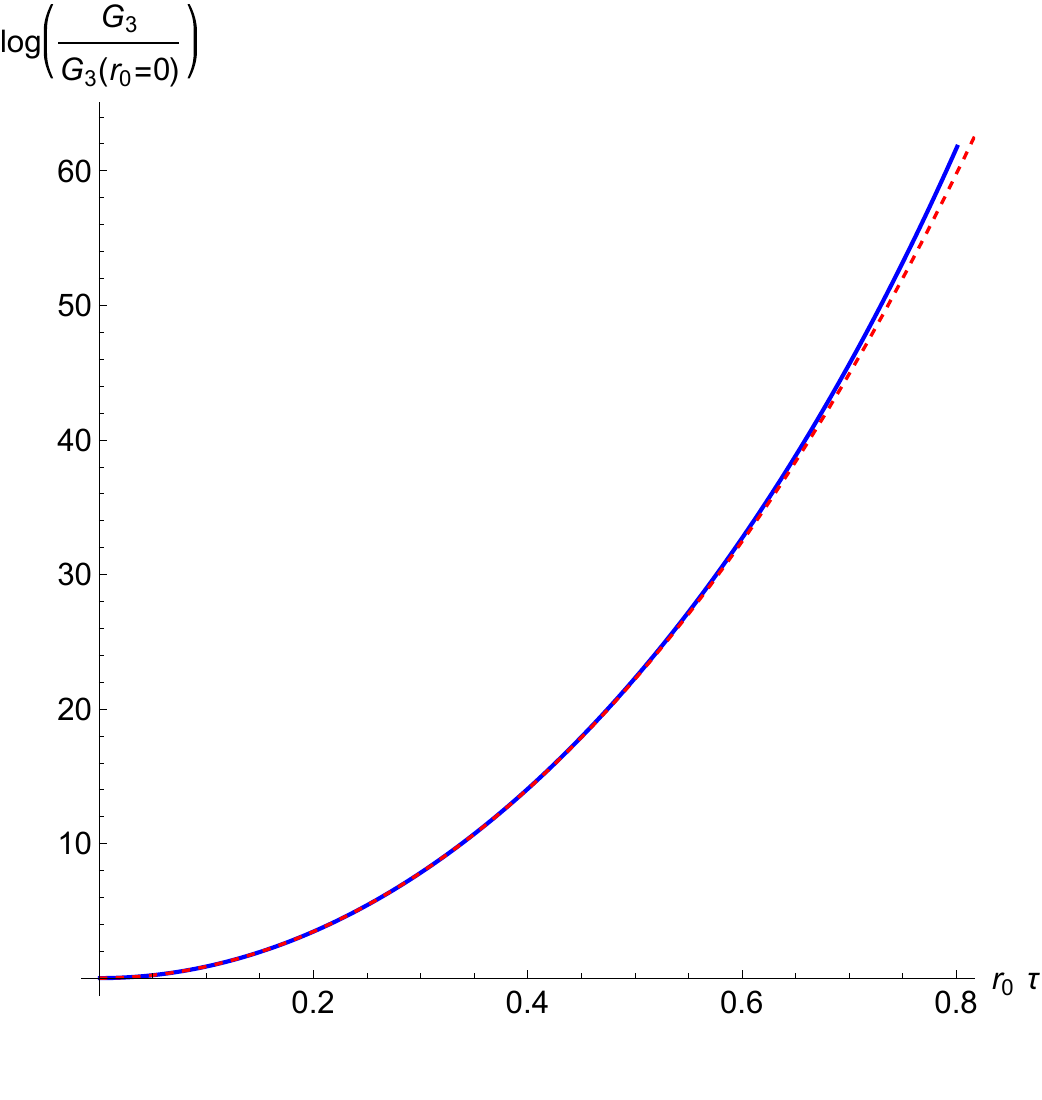}
     \caption{{\bf 1} in going and {\bf 2} returning geodesics for operators with conformal dimensions $\Delta_1= 155$, 
    $\Delta_2= 195$ and $\Delta_3= 200$ at the positions $\tau_1=0$, $\tau_2=1$ and $\tau_3=2$ of the boundary: 
    On the left part we plot the world-sheet of the three-point function. 
    The orange curve corresponds to $r_0=0.901$ and the blue (dashed) curve to $r_0=0.601$.
    The two straight lines determine the position of the singularity for each one of the two values of the parameter $r_0$, 
    i.e. $z_{\rm singular}=r_0^{-1}$. 
    On the right part, we plot the logarithm of the three point function for arbitrary value of the parameter $r_0$ 
    over the three point function for $r_0=0$ as a function of the dimensionless quantity $\tau T$. In the current case we
     have chosen a set-up for which $\tau_3=2 \, \tau_2$ and $\tau_2=1$. As a result, the parameter that changes in the 
     horizontal axis is the value of $r_0$. The blue curve is the result of the numerical 
    computation and for the red (dashed) curve we have used \eqref{3point-R-action-perturbative} 
    that contains the first and the second order corrections to the parameter $r_0$.
    } 
   \label{3point-R-12}
\end{figure}

According to the discussion at the end of section \ref{setup}, in order to determine the meeting point of the three-point function in the bulk, one needs to solve an algebraic system of 
five equations with five unknowns, i.e. $\tau_I$, $z_I$, $\mu_1$, $\mu_2$ \& $\mu_3$. 
Notice that the second equation in  \eqref{ti-zi} can be derived by taking into account that $S$ depends on 
$z_I$ both explicitly and implicitly through the dependence of $\mu_i$ on $z_I$.  
One can determine this dependence by differentiating \eqref{incoming-geodesic} with respect to $z_I$ 
in order to find ${d\mu_i/d z_I}$. 
This should then be plugged to the expression for the derivative of the action with respect to $z_I$
\begin{equation} \label{dszi}
\frac{dS_i}{dz_I} \, = \, 
\frac{\partial S_i}{\partial z_I} \,+ \, \frac{\partial S_i}{\partial \mu_i} \, 
\frac{\partial \mu_i}{\partial z_I} 
\quad  \Rightarrow \quad 
\frac{dS_i}{dz_I} \, = \, \Delta_i  \,  \frac{\sqrt{h(z_I) }}{z_I \, \left(1- r_0^2 \, z_I^2\right)}
\end{equation}
which in turn implies the second equation in \eqref{ti-zi}.
Solving numerically the system of the five equations we obtain the plots of the three-point correlator as a function of 
the deformation parameter $r_0$ that are depicted in figures  \ref{3point-R-21} and \ref{3point-R-12}. Below we summarize our findings.

On the left part of figure \ref{3point-R-21} we plot the worldlines corresponding to the three-point function 
with 2 incoming and 1 returning geodesics,  
that comprises of operators with conformal dimensions $\Delta_1= 95$, $\Delta_2= 135$ and $\Delta_3= 200$ 
at the positions $\tau_1=0$, $\tau_2=1$ and $\tau_3=2$ respectively. The two colors correspond to two different values for the 
parameter $r_0$. More specifically, blue corresponds to $r_0=0.509$ and orange to $r_0=0.709$ (notice that we work in units $R=1$). In the same plot 
we have also included the position of the singularity, i.e. the position where the distribution of D3-branes is situated
with a straight line. The singularity of the geometry sits at the bulk point $z=r_0^{-1}$ and is depicted by two lines one for each value of $r_0$. 
Notice that when the value of the parameter $r_0$ increases, the meeting point penetrates deeper in the bulk and 
the distance from the singularity (the straight line that appears in the plot) decreases. 
This suggests that if we keep increasing the value of $r_0$ the computation should collapse, since the world-sheet 
would approach (or even touch) the singularity.  In practice, in the example we present and in the ones we have studied, 
the numerical computation stops giving real solutions for the holographic coordinate $z$ already when there is considerable distance between the world-sheet of the three-point 
point function and the singularity, that is well before the world-lines touches the singularity. 
As discussed in the papers  \cite{Klose:2011rm,Minahan:2012fh} the based on geodesic arcs  derivation of the well known formula for the three-point functions is valid 
when a triangle inequality for the conformal dimensions of the operators is imposed. 
It seems that in the presence of the R-charge density this inequality needs to be modified and it is because of 
this modification that the calculation collapses, even before we approach the singularity. A second possibility is that another saddle point dominates after the critical point at which the solution becomes complex.

On the right part of figure \ref{3point-R-21} we plot the three point function for arbitrary values of the rotation 
parameter $r_0$ divided by the three point function for $r_0=0$ as a function of the dimensionless quantity $r_0 \tau$. 
We have chosen to place the operators at $\tau_1=0$, $\tau_2=1$ and $\tau_3=2$, and 
$\tau=t_3-t_2=t_2-t_1=1$ in the horizontal axis. 
The blue curve is the result of the numerical computation, while for the red curve we have used the result of the 
perturbative computation from equation \eqref{3point-R-action-perturbative} after fixing the corresponding values for $\alpha_1$,  
$\alpha_2$,  $\alpha_3$,  $\tau_1$, $\tau_2$ and $\tau_3$, namely  $\alpha_1=240$,  $\alpha_2=160$,  $\alpha_3=30$,
$\tau_1=0$, $\tau_2=1$ and $\tau_3=2$. The relation between $\alpha_1$,  $\alpha_2$,  $\alpha_3$ and the conformal 
dimensions $\Delta_1$,  $\Delta_2$,  $\Delta_3$ is given below in \eqref{alpha-definition}. For sufficiently small values of the 
the rotation parameter $r_0$ the perturbative solution is very accurate and captures the monotonically increasing behavior 
of the three-point function.

In figure \ref{3point-R-12} we plot the world-sheet (left part) and 
the three point function (right part) for operators that are described by 1 incoming and 2 returning geodesics, 
that comprises of operators with conformal dimensions  $\Delta_1= 155$, $\Delta_2= 195$ and $\Delta_3= 200$ 
at the positions $\tau_1=0$, $\tau_2=1$ and $\tau_3=2$. 
The behavior both for the world-sheet and for the value of the three-point function is similar to the one we have just described 
in figure \ref{3point-R-21}.

An important comment is in order: Notice that the analytic perturbative computation that will lead to equation 
\eqref{3point-R-action-perturbative} is performed for a three-point function with 2 in going and 1 returning 
geodesics. However, as can be seen from the right part of figure \ref{3point-R-12}, that relation describes also the 
result for small values of $r_0$ of a three-point function with 1 in going and 2 returning geodesics. 
Even if such a statement sounds plausible, it is highly non trivial to check it analytically since 
all the intermediate steps of the computation, namely the values of $\delta^{(20)} \tau$, $\delta^{(20)} z$ and 
$ \delta^{(20)} \mu_i $ for $i =1,2,3$, are different between a three-point function of 2 in going and 1 returning 
geodesics and a three-point function with 1 in going and 2 returning geodesics. 
However, we have checked analytically that despite the fact that the intermediate expressions are different the 
final result for the three-point function is universal in the sense that it does not depend on the precise nature 
of the three geodesics.


\paragraph{Perturbative computation}

\noindent

\noindent
To supplement the numerical computation we have already described, 
we consider the most generic three-point correlator and compute its value perturbatively around 
the $r_0=0$ case, which corresponds to the $AdS_5$ result \cite{Klose:2011rm,Minahan:2012fh}.  
For this reason the three operators with conformal dimensions $\Delta_1$, $\Delta_2$ and $\Delta_3$ 
are placed at the boundary points $(\tau, z)= (0, 0)$, $(\tau_2, 0)$ and $(\tau_3, 0)$ respectively.  

There are two possible configurations for the three-point function of unequal operators. We can either have one returning 
 and two in going geodesics or two returning  and one in going geodesic. Even if the equations that needs to be solved 
and the expression for the action that will be exponentiated are different, the final perturbative result is the same. 
For this reason, one can start with either configuration for the three geodesics. 

As in the numerical computation of the three-point function that we thoroughly described in the beginning of  
\ref{3point-Rcharge}, to determine the  three-point correlator we have to solve an algebraic system of 
five equations. These are: the two equations from  \eqref{ti-zi} and the equations for the three geodesics arcs from the boundary points 
until the meeting point in the bulk. 
To solve the equations (for one returning and two in going geodesics) we expand around  the zero R-charge  
solution in the following way 
\begin{eqnarray} \label{R-charge-expansion-ansatz}
&& \tau_{I} = \tau_0 + r_0^2 \, \delta^{(20)}  \tau + r_0^4 \, \delta^{(40)} \tau \, ,  \quad 
z_{I} = z_0 +  r_0^2 \, \delta^{(20)} z + r_0^4 \, \delta^{(40)} z 
 \nonumber \\[5pt]
 &&   \quad \& \quad  
 \mu_i \, = \, \mu_{i}^0  + r_0^2 \, \delta^{(20)} \mu_i +  r_0^4 \, \delta^{(40)} \mu_i  \quad {\rm for} \quad  i =1,2,3 \, . 
\end{eqnarray}
We use two numbers to characterize the order of the perturbation. The first one is  the power of the parameter $r_0$
and the second is the power of the parameter $\eta$, that is related to the temperature. In the current section the value of this 
second number will always be zero. The reason for this notation will become evident when we 
have to combine the two deformations.
The coordinates of the meeting point at zero temperature  and zero R-charge are given by \cite{Klose:2011rm,Minahan:2012fh} 
\begin{eqnarray} \label{KM-tau-z}
&& \tau_0  \, = \,  \frac{ \alpha_1 \, \tau_2 \, \tau_3 \, \left(\alpha_2 \, \tau_2+ \alpha_3 \, \tau_3\right)}{\alpha_1 \, \alpha_2 \, \tau_2^2
+ \alpha_2 \, \alpha_3 \, \tau_3^2 +  \alpha_2 \, \alpha_3 \, \left(\tau_3 -\tau_2\right)^2} \nonumber \\[5pt] 
&& z_0  \, = \,  \frac{ \sqrt{\alpha_1 \, \alpha_2 \, \alpha_3 \, \left(\alpha_1 + \alpha_2 + \alpha_3\right)} \, 
\left(\tau_3 - \tau_2 \right) \, \tau_2 \, \tau_3}{\alpha_1 \, \alpha_2 \, \tau_2^2
+ \alpha_2 \, \alpha_3 \, \tau_3^2 +  \alpha_2 \, \alpha_3 \, \left(\tau_3 -\tau_2\right)^2} \quad {\rm for } \quad \tau_3>\tau_2
\end{eqnarray}
where the constants $\alpha_1$, $\alpha_2$ and $\alpha_3$ are expressed in terms of the conformal dimensions
\begin{equation} \label{alpha-definition}
\alpha_1\, = \, \Delta_2+ \Delta_3 - \Delta_1\, , \quad 
\alpha_2\, = \, \Delta_3+ \Delta_1 - \Delta_2 \quad \& \quad 
\alpha_3\, = \, \Delta_1+ \Delta_2 - \Delta_3
\end{equation}
and the values for the constants $\mu_{1}^0$, $\mu_{2}^0$ and $\mu_{3}^0$ are given by
\begin{eqnarray}  \label{KM-mu}
&&  \mu_{1}^0 =   \frac{2}{\alpha_2 + \alpha_3} \, \left[ \frac{\alpha_2}{\tau_3} + \frac{\alpha_3}{\tau_2} \right] \, , \quad 
\mu_{2}^0 = -\,  \frac{2}{\alpha_3 + \alpha_1} \, \left[ \frac{\alpha_1}{\tau_2 - \tau_3} + \frac{\alpha_3}{\tau_2} \right] 
\nonumber \\[5pt]
&& \quad \quad \quad \quad \& \quad \quad \quad 
\mu_{3}^0  =   \frac{2}{\alpha_1 + \alpha_2} \, \left[ \frac{\alpha_1}{\tau_2 - \tau_3} - \frac{\alpha_2}{\tau_3} \right] \, . 
\end{eqnarray}
Solving the system of the five equations at fourth order in $r_0$, we determine the values of the ten unknowns that appear 
in \eqref{R-charge-expansion-ansatz}. The coefficients at second order in $r_0$ are simple and we list them in this 
subsection, while those at fourth order are lengthier and for this reason they will not be presented. However, it is 
straightforward to calculate them.

For $ \delta^{(20)} \tau $ the result is
\begin{eqnarray} \label{3point-R-delta-tau}
&& \delta^{(20)} \tau = -  \frac{1}{6\, \Theta^3 }
\alpha_1 \, \alpha_2 \, \alpha_3 \, \tau_2^2 \, \tau_3^2 \, \left(\tau_3-\tau_2\right)^2    
\Big[\alpha_2  \, \tau_2^3  (\alpha_1+ \alpha_3) (\alpha_1 +2 \alpha_2+\alpha_3)
 \\[5pt]
 && - \, 3 \, \tau_2 \, \tau_3 (\alpha_1+ \alpha_2) (\alpha_1+ \alpha_3) (\alpha_2 \,\tau_2 +\alpha_3 \,\tau_3)
 +\alpha_3 \,\tau_3^3 (\alpha_1+ \alpha_2) (\alpha_1+\alpha_2+2 \,\alpha_3) \Big]
 \nonumber
\end{eqnarray}
where the constant $\Theta$ is defined as follows   
\begin{equation} \label{3point-R-Theta}
\Theta \,=\, \alpha_1 \, \alpha_2 \, \tau_2^2
+ \alpha_1 \, \alpha_3 \, \tau_3^2 +  \alpha_2 \, \alpha_3 \, \left(\tau_3 -\tau_2\right)^2 \, . 
\end{equation}
For $\delta^{(20)}z $ the result is
\begin{eqnarray} \label{3point-R-delta-z}
&&\delta^{(20)} z = \frac{z_0}{6\, \Theta^2} \,\tau_2 \, \tau_3 \left(\tau_3-\tau_2\right) 
\Big[ -2\, \alpha_1 \, \alpha_2 \, \alpha_3 \left(\alpha_1+\alpha_2+\alpha_3\right) \,\tau_2 \, \tau_3 \left(\tau_2-\tau_3\right)   
\\[5pt]
&&  \qquad \qquad 
 -  \, \alpha_1^2 \, \alpha_2^2 \, \tau_2^3  +  \, \alpha_1^2 \, \alpha_3^2 \, \tau_3^3-
  \alpha_2^2 \, \alpha_3^2 \, \left(\tau_3-\tau_2\right)^3 \Big]
 \nonumber
\end{eqnarray}
while for $ \delta^{(20)} \mu_1 $, $ \delta^{(20)} \mu_2$ and 
$ \delta^{(20)}  \mu_3$ we obtain
\begin{equation}  \label{R-charge-delta-mu1-mu2-mu3}
\delta^{(20)} \mu_1 = -\, \frac{1}{3} \, \frac{\alpha_3 \, \tau_2 +\alpha_2 \, \tau_3}{\alpha_2+\alpha_3} \, ,\quad 
\delta^{(20)} \mu_2 = \frac{1}{3} \left[\tau_2-\frac{\alpha_1 \, \tau_3}{\alpha_1+ \alpha_3} \right] \, ,
\quad 
\delta^{(20)}\mu_3 =\frac{1}{3} \left[\tau_3-\frac{\alpha_1 \, \tau_2}{\alpha_1+ \alpha_2} \right] \, . 
\end{equation}
By inspecting \eqref{3point-R-delta-z} one sees that $\delta^{(20)} z\sim z_0$. This proves that in the case of extremal correlators $\delta^{(20)} z=0$ since $z_0=0$. The same holds true for the  second correction $\delta^{(40)} z=0$, too. As a result we have shown that the extremal three-point correlator factorises into the product of two two-point correlators as it happens in the $AdS_5\times S^5$ case.

The total action that will be exponentiated comprises of one returning and two in going pieces, i.e. 
\begin{equation}  \label{R-action-perturbative_v0}
S_{\rm total} = 2 \, S_{\rm max}^1-S_{\rm in}^1+ S_{\rm in}^2 + S_{\rm in}^3 \, . 
\end{equation}
Substituting the expansion ansatz of \eqref{R-charge-expansion-ansatz}, expanding at fourth order in $r_0$ 
and plugging in the expressions for the coefficients that appear in this subsection, 
we arrive to the following result for the three-point function
\begin{equation} \label{3point-R-action-perturbative}
S^{\rm R}_{\rm total} \, = \,S_{\rm KM} - \frac{r_0^2}{12} \, \delta S^{(20)} + 
\frac{r_0^4}{360 \, \Theta} \, \delta S^{(40)}
\end{equation}
where $S_{\rm KM}$ is the value of the three-point function in $AdS_5$ that can be found in 
  \cite{Klose:2011rm,Minahan:2012fh}
\begin{equation}
S_{\rm KM} \, = \, -\,  \log \Bigg[ \frac{{\cal C}_{{\cal O}_{\Delta_1}{\cal O}_{\Delta_2} {\cal O}_{\Delta_3} }}{\tau_2^{\alpha_3} \tau_3^{\alpha_2} 
\left(\tau_3-\tau_2\right)^{\alpha_1} }\Bigg]  
\end{equation}
and the expression for the coefficient  ${\cal C}_{{\cal O}_{\Delta_1}{\cal O}_{\Delta_2} {\cal O}_{\Delta_3} } $ is given by
\begin{equation}
 {\cal C}_{{\cal O}_{\Delta_1}{\cal O}_{\Delta_2} {\cal O}_{\Delta_3} }  =
 \Bigg[\frac{\alpha_1^{\alpha_1}\, \alpha_2^{\alpha_2}\, \alpha_3^{\alpha_3} \, 
 \left(\alpha_1+\alpha_2+\alpha_3\right)^{\alpha_1+\alpha_2+\alpha_3}}{ \left(\alpha_1+\alpha_2\right)^{\alpha_1+\alpha_2}\, \left(\alpha_2+\alpha_3\right)^{\alpha_2+\alpha_3}\, \left(\alpha_3+\alpha_1\right)^{\alpha_3+\alpha_1}}\Bigg]^{1/2} \, . 
\end{equation}
The expression for the correction of the three point function at second order in $r_0$ is
\begin{equation} \label{3point-R-action-coeff-2}
\delta S^{(20)} \, = \, \alpha_3 \, \tau_2^2 +  \alpha_2 \, \tau_3^2 + \alpha_1 \, \left(\tau_3-\tau_2\right)^{2} 
\end{equation}
while the correction at fourth order in $r_0$ is
\begin{eqnarray} \label{3point-R-action-coeff-4}
&& \delta S^{(40)}\, = -2 \, \alpha _1^2 \left(\tau _2-\tau _3\right)^4 
\left(\alpha _2 \, \tau _2^2+\alpha _3 \, \tau _3^2\right)- 2 \, \alpha _2 \, \alpha _3 
\left(\tau _2-\tau _3\right)^2 \left(\alpha _3 \, \tau _2^4+\alpha _2 \, \tau _3^4\right) 
\nonumber \\[5pt]
&&
+ \alpha _1 \Big[-4 \, \alpha _2 \, \alpha _3 \, \tau _2^6+12 \, \alpha _2 \, \alpha _3 \, \tau _3 \, \tau _2^5 - 
\alpha _3 \left(15 \, \alpha _2+2 \, \alpha _3\right) \tau _3^2 \tau _2^4 +10 \, \alpha _2 \, \alpha _3 \, \tau _3^3 \, \tau _2^3 
\nonumber \\[5pt]
&& -\alpha _2 \left(2 \, \alpha _2+15 \, \alpha _3\right) \tau _3^4 \, \tau _2^2+12 \, \alpha _2 \, \alpha _3 \, \tau _3^5 \, \tau _2
-4 \, \alpha _2 \, \alpha _3 \, \tau _3^6\Big] \, . 
\end{eqnarray}

On the right part of figure \ref{3point-R-12} we put a challenge on the perturbative result of the 
three-point function in \eqref{3point-R-action-perturbative}. We compare the numerical with the 
perturbative computation for values of the dimensionless parameter $r_0 \tau$ between 0 and 0.9. 
We find an amazing agreement and the two results begin to diverge after $r_0 \tau>0.8$. 
This is very promising for two reasons: From one side it confirms the perturbative computation but
most importantly it is telling us that we can trust the perturbative analysis even when the perturbative parameter is not 
very small. It is clear that if we could add one more term in the perturbative expansion 
\eqref{3point-R-action-perturbative} (i.e. the sixth order term in $r_0$) the numerical and perturbative results 
would coincide for an even larger range of values of $r_0$.



\subsection{Finite temperature}
\label{3point-Temperarure}

The next step in the detailed exploration of three-point correlation functions is to consider a background that only contains 
temperature. The computation in this framework has been examined in 
\cite{Rodriguez-Gomez:2021mkk}, but here we will extend the discussion to include a case that is not covered in \cite{Rodriguez-Gomez:2021mkk}, namely the case in which the boundary 
distance between the correlators is spatial. Moreover, we will obtain perturbative analytical expressions for the 
three-point correlators, when the boundary distance is both temporal and spatial. 


\subsubsection{Temporal boundary distance}

The expressions for the temporal boundary distance and the action can be found either from equation 
\eqref{2point-tdot-positive-v1} or from equations  \eqref{2point-tdot-negative-v1} and \eqref{2point-action-negative-v1}.
Lets emphasize here that in the absence of the R-charge those expressions become identical. As in the case we examined
in subsection \ref{3point-Rcharge} those integrals can be calculated analytically and the results are the following
 \begin{equation}
 \label{T-action-integral-mu}
- \, \frac{2 \, S}{\Delta} = \log \left[\epsilon ^2 \, 
\frac{2 -\mu ^2\, z^2 +2  \sqrt{h(z)}}{4 \, z^2}\right]
\quad {\rm with} \quad h(z) = 1 - \eta^4 \, z^4 - \mu ^2 \, \,z^2 
\end{equation}
and
 \begin{eqnarray}  \label{T-t-t1-mu}
 \tau-\tau_1 &=& - \, \frac{1}{4 \, \eta} \, \log \left[\frac{\mu ^2-  \left(\mu ^2-2 \, \eta^2\right) \eta ^2 \, z^2 + 2 \, \eta \,  \mu  
 \sqrt{h(z)} + 2 \,  \eta ^2}{  \left(1+ \eta ^2 \, z^2 \right) 
 \left(\mu ^2 + 2 \, \eta \,  \mu + 2 \, \eta^2\right)}\right] 
  \nonumber  \\[7pt]
   && + \,  \frac{i}{4 \, \eta} \, \log \left[\frac{\mu ^2 + \left(\mu ^2 + 2 \, \eta^2\right) \eta ^2 \, z^2 +  2\, i  \, \eta \,  \mu  
 \sqrt{h(z)} - 2 \,  \eta ^2}{  \left(1- \eta ^2 \, z^2 \right) 
 \left(\mu ^2 +2 \, i \, \eta \,  \mu - 2 \, \eta^2\right)}\right] \, . 
\end{eqnarray}
To determine the meeting point needed for the calculation of the three-point function, we have to specify the conformal dimensions of the three 
operators and their boundary location. To facilitate the comparison with the results that were presented in the previous 
subsection, we choose the same values both for the conformal dimensions and the boundary points. 

\begin{figure}[ht] 
   \centering
   \includegraphics[width=7.5cm]{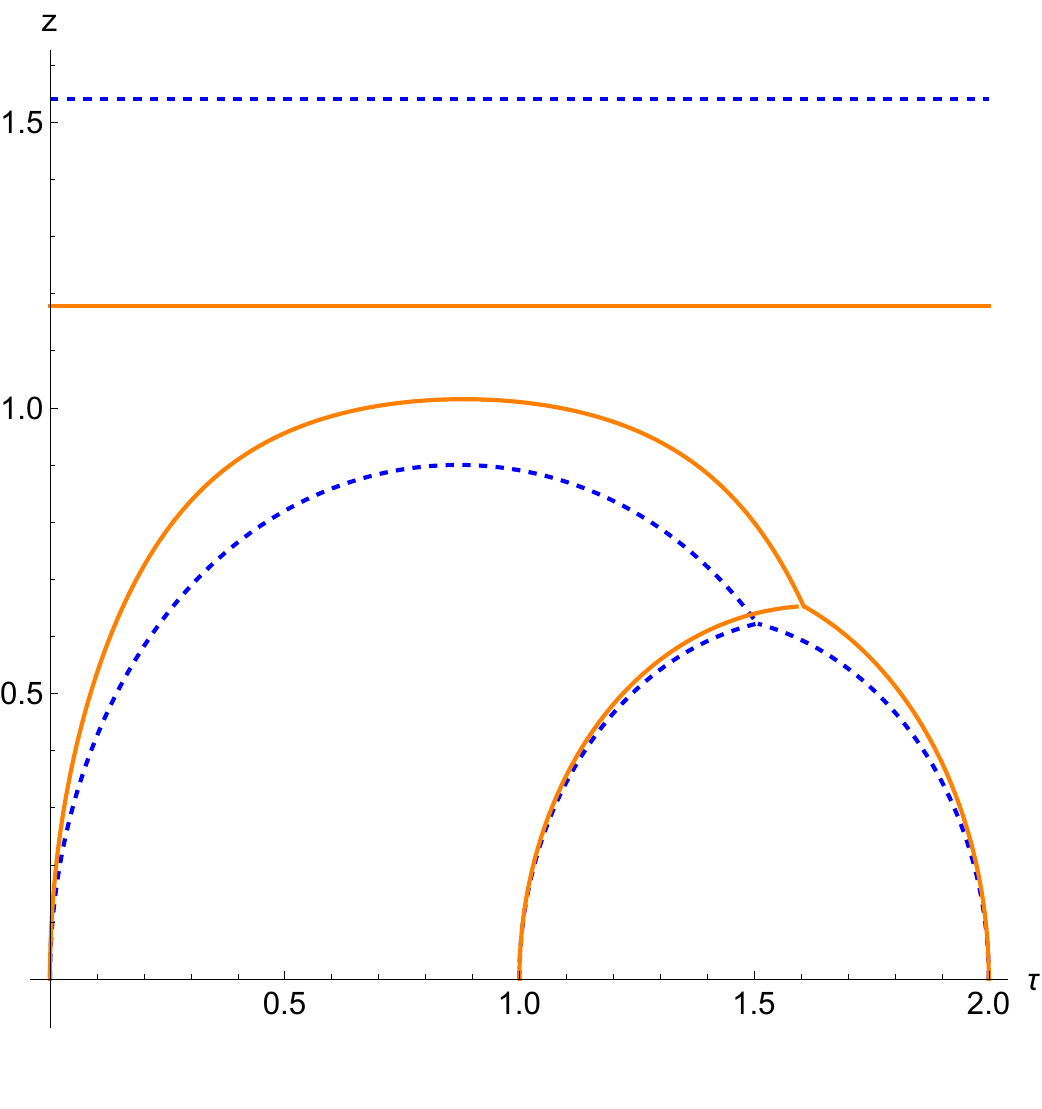}
   \includegraphics[width=7.5cm]{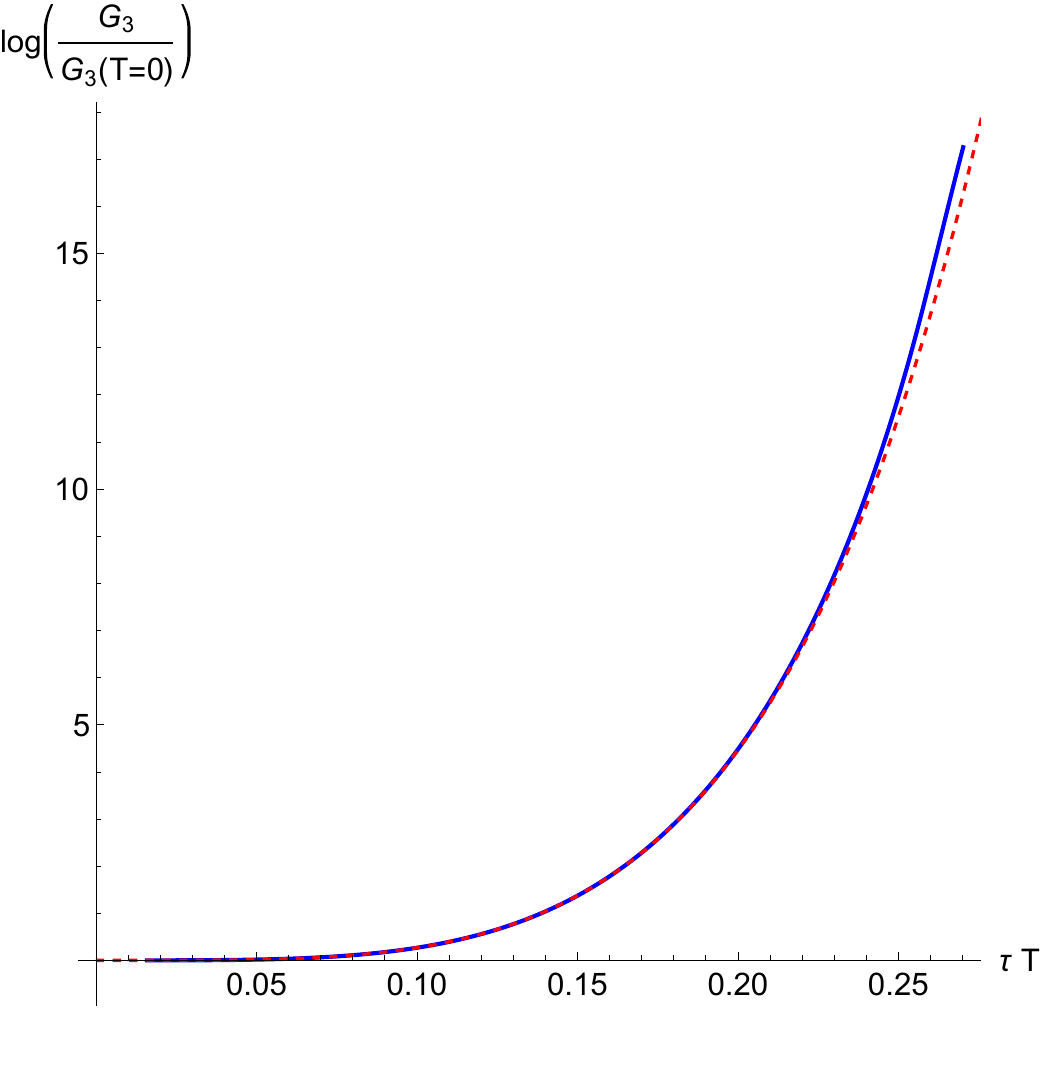}
    \caption{{\bf 2} in going and {\bf 1} returning geodesics for operators with conformal dimensions $\Delta_1= 95$, 
    $\Delta_2= 135$ and $\Delta_3= 200$ at the (temporal) positions $\tau_1=0$, $\tau_2=1$ and $\tau_3=2$ of the boundary: 
    On the left part we plot the world-sheet of the three-point function. 
    The orange curve corresponds to $\eta=0.849$  (or else $T=0.270$) and the blue (dashed) curve to $\eta=0.649$ 
    (or else $T=0.207$).
    The two straight lines determine the position of the horizon for each one of the two values of the parameter $\eta$, 
    i.e. $z_{h}=\eta^{-1}$. 
    On the right part, we plot the logarithm of the three point function for arbitrary temperature over the three point 
    function for zero temperature as a function of the dimensionless quantity $\tau T$. In the current case we
     have chosen a set-up for which $\tau_3=2 \, \tau_2$ and $\tau_2=1$. As a result, the parameter that changes in the 
     horizontal axis is the value of $T$. The blue curve is the result of the numerical 
    computation and for the red (dashed) curve we have used \eqref{T-temporal-3point-perturbative_v2}
    that contains the first and the second order corrections to the parameter $\eta$ (or else $T$).}
   \label{T-temporal-3point-21}
\end{figure}

On the left part of figure \ref{T-temporal-3point-21} we plot the world-sheet of the three-point function with 2 incoming and 1 returning geodesics, that comprises of operators with conformal dimensions $\Delta_1 =95$ , $\Delta_2 =135$ 
and $\Delta_3 =200$ at the positions $\tau_1 =0$, $\tau_2=1$ and $\tau_3 =2$. 
The two colors correspond to two different values for the parameter $\eta$ (or equivalently of the temperature). 
More specifically, blue corresponds to $\eta=0.649$ and orange to $\eta=0.849$ (notice that as in subsection 
\ref{3point-Rcharge} we work in units R = 1). 
In the same plot we have also included the position of the horizon  
with a straight line at the bulk point $z=\eta^{-1}$ (two lines for each one of the values of $\eta$).
The main conclusion from the comparison of the blue and the orange curves is identical to one we reached 
from the study of figures \ref{3point-R-21} and \ref{3point-R-12}.
As the value of the deformation parameter $\eta$ increases, the meeting point penetrates deeper in the bulk and 
the distance from the horizon (the straight line that appears in the plot) decreases. 
If we keep increasing the value of $\eta$, for the current choice of conformal dimensions, the world-sheet 
touches the horizon and afterwards the computation collapses. However, for different choices of conformal dimensions, 
the numerical computation stops giving real solutions when there is considerable distance between the world-sheet of the three-point 
point function and the horizon. 

On the right part of figure \ref{T-temporal-3point-21}, we plot the logarithm of the three point function for arbitrary 
temperature over the three point function for zero temperature as a function of the dimensionless quantity $\tau T$. 
Choosing a set-up in which $\tau_3=2 \, \tau_2$ and $\tau_2=1$, the parameter that effectively changes along the 
horizontal axis is the value of $T$. 
The blue curve is the result of the numerical computation and for the red (dashed) curve we have used 
\eqref{T-temporal-3point-perturbative_v2} that contains the first and the second order corrections to the parameter 
$\eta$ (or else $T$). The monotonic behavior of the value of the three-point function is similar to the one we observed in 
figures \ref{3point-R-21} and \ref{3point-R-12} of the previous subsection.

\paragraph{Perturbative computation}

\noindent

\noindent
Following the discussion of the previous subsection, we will consider again the generic expressions for the five 
equations that determine the meeting point for the three-point correlator and expand around the zero temperature 
result of \cite{Klose:2011rm,Minahan:2012fh}. The expansion ansatz will be  
\begin{eqnarray} \label{T-expansion-ansatz}
&& \tau_{I} \, = \, \tau_0 \,+ \, \eta^4 \, \delta^{(04)} \tau + \, \eta^8 \, \delta^{(08)} \tau \, ,  \quad 
z_{I} \, = \, z_0\,+ \, \eta^4 \, \delta^{(04)} z + \, \eta^8 \, \delta^{(08)} z 
 \nonumber \\[5pt]
 &&   \quad \& \quad  \mu_i \, = \, \mu_{i  0} \,  + \, \eta^4 \, \delta^{(04)} \mu_i + \, \eta^8 \, \delta^{(08)} \mu_i  \, ,  
 \quad i =1,2,3
\end{eqnarray}
where the first term inside each parenthesis (i.e. index) is zero, since for the computations of this subsection $r_0=0$. 
The coordinates of the meeting point at zero temperature are given by the expressions that we presented in the previous
subsection, while the coefficients at fourth order in $\eta$ are analytic but very lengthy and for this reason 
they will not be presented here. However, it is straightforward to calculate them. 

Substituting the expansion ansatz of \eqref{T-expansion-ansatz}, expanding at fourth order in $\eta$ 
and plugging in the complicated expressions for the coefficients, 
we arrive to the following expression for the three-point function
\begin{equation} \label{T-temporal-3point-perturbative}
S_{\rm total}  \, = \,S_{\rm KM} - \frac{\eta^4}{80 \, \Theta} \, \delta S^{(04)}
\end{equation}
where the expression for  $\delta S^{(04)}$ reads
\begin{eqnarray} \label{T-temporal-action_first_correction_general}
&&  \delta S^{(04)} \, =  \alpha_1^2 
\left(\tau_2 - \tau_3\right)^4 \left(\alpha_2 \, \tau_2^2 +\alpha_3 \, \tau_3^2\right) 
+ \, \alpha_2 \, \alpha_3 \left(\tau_2 - \tau_3\right)^2 \left(\alpha_3 \, \tau_2^4 +\alpha_2 \, \tau_3^4\right) + 
\alpha_1  \big(\alpha_3^2 \, \tau_2^4 \, \tau_3^2  
\nonumber \\[5pt]
&& +\alpha_2^2 \, \tau_3^4 \, \tau_2^2 \big) 
 + \, \alpha_1 \, \alpha_2 \, \alpha_3 \, \left( 2\, \tau_2^6 - 6 \, \tau_2^5 \, \tau_3 +  5 \, \tau_2^4 \, \tau_3^2 + 
5 \, \tau_3^4 \, \tau_2^2 - 6 \, \tau_3^5 \, \tau_2 + 2\, \tau_3^6\right) \, . 
\end{eqnarray}
Trying to go beyond the fourth order in the perturbative expansion of \eqref{T-temporal-3point-perturbative}, 
gives extremely complicated expressions for any of the ingredients of the solutions. 
However, if the boundary distance between the second and the third operator is fixed according to the relation
$\tau_3 = 2 \, \tau_2$, then the calculations simplify and it is possible to expand the action at eighth  order in $\eta$. 
The result we obtain  is the following 
\begin{equation} \label{T-temporal-3point-perturbative_v2}
S_{\rm total}^{\, \tau_3\, = \, 2 \, \tau_2}  \, = \,S_{\rm KM}^{\, \tau_3\, = \, 2 \, \tau_2} - \frac{\eta^4\, \tau_2^4}{80\, \Theta} \,
 \delta S^{(04)}_{\, \tau_3\, = \, 2 \, \tau_2} -
 \frac{\eta^8\, \tau_2^8}{28800\, \Theta^4}  \, \delta S^{(08)}_{\, \tau_3\, = \, 2 \, \tau_2} 
\end{equation}
where $ \delta S^{(04)}_{\, \tau_3\, = \, 2 \, \tau_2}$ can be obtained from the general expression 
\eqref{T-temporal-action_first_correction_general} by setting $\tau_3\, = \, 2 \, \tau_2$
\begin{equation} \label{T-temporal-3point-first-correction}
 \delta S^{(04)}_{\, \tau_3\, = \, 2 \, \tau_2}\, = \,\alpha_1^2 \left(\alpha_2 +4 \, \alpha_3 \right) + 
16 \, \alpha_2^2 \left(\alpha_1 + \alpha_3 \right) + 
\alpha_3^2 \left(4 \, \alpha_1 + \alpha_2 \right) + 
26\, \alpha_1 \, \alpha_2 \, \alpha_3
\end{equation}
and the eighth order correction $ \delta S^{(08)}_{\, \tau_3\, = \, 2 \, \tau_2}$ is
\begin{eqnarray}  \label{T-temporal-3point-second-correction}
&& \delta S^{(08)}_{\, \tau_3\, = \, 2 \, \tau_2}= 11 \, \alpha_1^5 \left(\alpha_2 +4 \alpha_3\right)^4 
+ \, \alpha_1 \,  \alpha_2^2 \, \alpha_3^4 \, 
\left(86624 \, \alpha_1^2 + 82880 \, \alpha_1 \, \alpha_2 + 32631 \, \alpha_2^2 \right)  
\nonumber \\[5pt] 
&&
+ \, 2816 \, \alpha_2^5 \left(\alpha_1 + \alpha_3\right)^4 
 \,+ \,  2 \,\alpha_1^2 \, \alpha_2^3 \, \alpha_3^2 \left(33319 \, \alpha_1 \, \alpha_2 + 63504 \, \alpha_1 \, \alpha_3 +
33319 \, \alpha_2  \, \alpha_3 \right) 
\\[5pt] 
&&
+\, \alpha _1^4 \, \alpha_2 \,  \alpha_3 
\left(32631 \, \alpha_2^3+ 82880 \, \alpha_2^2 \, \alpha_3  + 86624 \, \alpha_2 \, \alpha_3^2+ 43008 \, \alpha_3^3\right) + 
11 \,\alpha_3^5 \left(4 \,\alpha_1+ \alpha_2\right)^4 \, . 
\nonumber
\end{eqnarray}
The perturbative result for the three-point function in \eqref{T-temporal-3point-perturbative_v2} can be compared with the 
result of the symmetric three-point function (i.e. $\Delta_1=\Delta_3=\Delta$) of \cite{Rodriguez-Gomez:2021mkk}. More 
specifically setting $\alpha_1=\alpha_3=\Delta_2$ and $\alpha_2=2\, \Delta - \Delta_2$ we arrive to following expression for
the action
\begin{eqnarray}
S_{\rm sym} &= & S_{\rm KM}^{\rm sym} - \frac{32 \Delta ^2-18 \Delta _2 \Delta +3 \Delta _2^2}{40 
\left(2 \Delta +\Delta_2\right)} \, \eta^4\, \tau_2^4 
\nonumber \\ 
& -&\frac{22528 \Delta ^4-17944 \Delta _2 \Delta ^3+2652 \Delta _2^2 \Delta ^2+974 
\Delta_2^3 \Delta -227 \Delta _2^4}{14400 \left(2 \Delta +\Delta _2\right){}^3}\, \eta^8\, \tau_2^8 \, . 
\end{eqnarray}
Let us mention that the first line of the equation above is identical with equation (4.11) of  \cite{Rodriguez-Gomez:2021mkk} while the second line is a new result.


\subsubsection{Spatial boundary distance}

The boundary distance between the operators in a three-point correlator can be either temporal or spatial. In the analysis we 
presented in section \ref{setup} and in order to simplify notation we have restricted to the temporal case. However, as can 
seen from the metric \eqref{metric-general-v1}, in the presence of temperature there is a difference between the two cases. 
Starting from this observation, in this subsection we will analyze three-point correlators in a background that  contains only
temperature with the boundary distance of the operators being spatial. 

The expressions for $\dot x$ and the action are 
\begin{equation} \label{T-spatial-xdot-action}
\dot x = \frac{\nu \, z}{\sqrt{\big.h(z)\left(1- \eta^4 \, z^4\right)}}
\quad \& \quad 
\frac{S}{\Delta} = \int \frac{dz}{z} \frac{1}{\sqrt{\big.h(z)\left(1- \eta^4 \, z^4 \right)}}
\quad {\rm with} \quad h(z) =   1- \nu^2 \,  z^2
\end{equation}
and analytic expressions for those integrals can be found in \cite{Georgiou:2022ekc}. 
The numerical solution of the algebraic system of equations that determine the meeting point of the three-point correlator 
in the bulk proceeds as in the previous cases.

\begin{figure}[ht] 
   \centering
   \includegraphics[width=7.5cm]{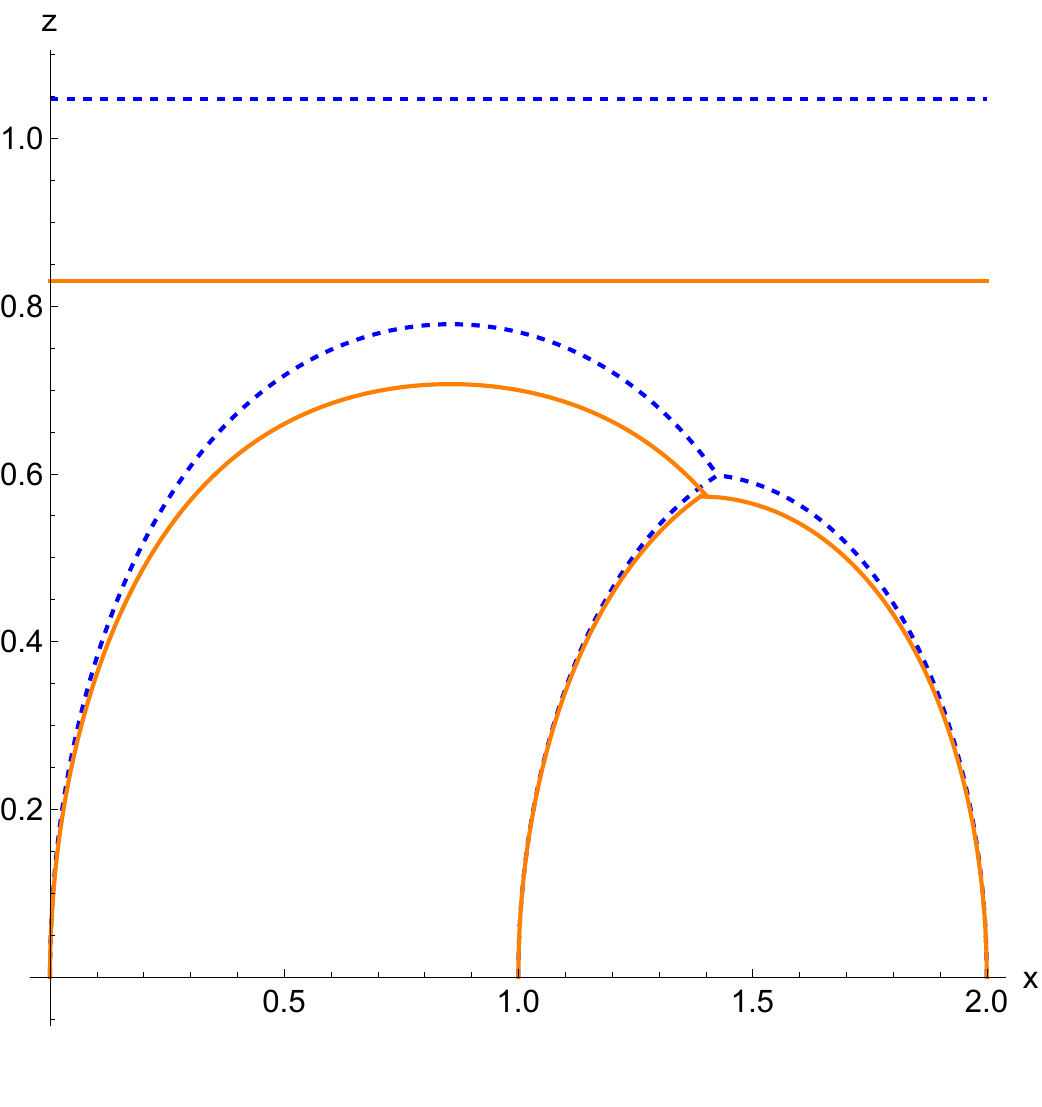}
   \includegraphics[width=7.5cm]{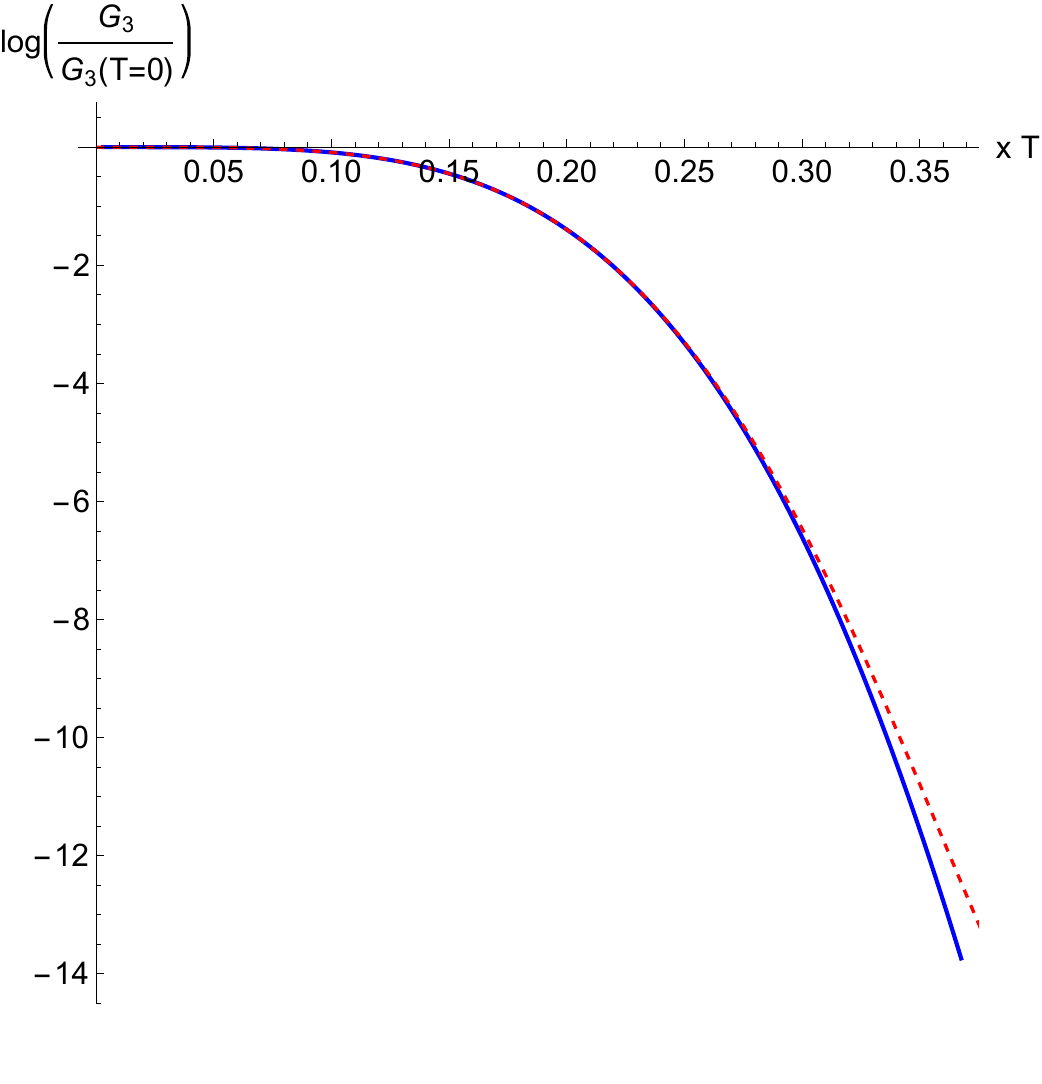}
    \caption{{\bf 2} ingoing and {\bf 1} returning geodesics for operators with conformal dimensions $\Delta_1= 95$, 
    $\Delta_2= 135$ and $\Delta_3= 200$ at the (spatial) positions $x_1=0$, $x_2=1$ and $x_3=2$ of the boundary: 
    On the left part we plot the worldsheet of the three-point function. 
    The orange curve corresponds to $\eta=1.205$  (or else $T=0.384$) and the blue (dashed) curve to $\eta=0.955$ 
    (or else $T=0.304$).
    The two straight lines determine the position of the horizon for each one of the two values of the parameter $\eta$, 
    i.e. $z_{h}=\eta^{-1}$. 
    On the right part, we plot the logarithm of the three point function for arbitrary temperature over the three point 
    function for zero temperature as a function of the dimensionless quantity $x T$. In the current case we
     have chosen a set-up for which $x_3=2 \, x_2$ and $x_2=1$. As a result, the parameter that changes in the 
     horizontal axis is the value of $T$. The blue curve is the result of the numerical 
    computation and for the red (dashed) curve we have used \eqref{T-spatial-3point-perturbative_v2}
    that contains the first and the second order corrections to the parameter $\eta$ (or else $T$).}
   \label{T-spatial-3point-21}
\end{figure}

On the left part of figure \ref{T-spatial-3point-21} we plot the world-sheet of the three-point function with 2 incoming and 1 returning geodesics, that comprises of operators with conformal dimensions $\Delta_1 =95$ , $\Delta_2 =135$ 
and $\Delta_3 =200$ at the spatial positions $x_1 =0$, $x_2=1$ and $x_3 =2$. 
The blue curve corresponds to $\eta=0.955$ and the orange one to $\eta=1.205$. 
The position of the horizon is depicted with a straight line at the bulk point $z=\eta^{-1}$ 
(two lines for each one of the values of $\eta$).
As the value of the deformation parameter $\eta$ increases, the meeting point moves closer to the boundary (i.e. the area 
of the world-sheet decreases) and the distance from the horizon (i.e. the straight line that appears in the plot) decreases. 
This behavior is different from the one we observed in cases with temporal boundary distance between the operators, in 
the previous subsection. 
If we keep increasing the value of $\eta$, for the current choice of conformal dimensions, the computation stops giving real solutions 
well before the world-sheet touches the horizon. 

On the right part of figure \ref{T-spatial-3point-21}, we plot the logarithm of the three point function for arbitrary 
temperature over the three point function for zero temperature as a function of the dimensionless quantity $x T$. 
Similarly to previous plots, the operators are placed at the boundary points $x_3=2 \, x_2$ and $x_2=1$ and as a result 
the parameter that effectively changes along the horizontal axis is the value of the temperature. 
The blue curve is the result of the numerical computation and for the red (dashed) curve we have used 
\eqref{T-spatial-3point-perturbative_v2} that contains the first and the second order corrections to the parameter 
$\eta$ (or else $T$). In agreement with the observation that the worldsheet shrinks as the temperature increases, the 
three-point function decreases as the dimensionless quantity $x T$ increases.

\noindent

\paragraph{Perturbative computation}

\noindent

\noindent
The numerical computation can be supplemented with a perturbative expansion around the zero temperature result.
It should be emphasized that in all the cases we have covered so far the starting point for every numerical 
construction of the three-point workdsheet is $AdS_5$ result of \cite{Klose:2011rm,Minahan:2012fh}. 
The expansion ansatz is the following
\begin{eqnarray} \label{T-spatial-expansion-ansatz}
&& x_{I} \, = \, \tau_0 \,+ \, \eta^4 \, \delta^{(04)} x + \, \eta^8 \, \delta^{(08)} x \, ,  \quad 
z_{I} \, = \, z_0\,+ \, \eta^4 \, \delta^{(04)} z + \, \eta^8 \, \delta^{(08)} z 
 \nonumber \\[5pt]
 &&   \quad \& \quad  \nu_i \, = \, \mu_{i  0} \,  + \, \eta^4 \, \delta^{(04)} \nu_i + \, \eta^8 \, \delta^{(08)} \nu_i  \, ,  
 \quad i =1,2,3 \, . 
\end{eqnarray}
While the coefficients at fourth order in $\eta$ are analytic but very lengthy to present them here, 
the coefficients at eighth order in $\eta$ is almost impossible to obtain them in closed form for generic values of the conformal
dimensions and boundary positions. 
Substituting the expansion ansatz of \eqref{T-spatial-expansion-ansatz}, expanding at fourth order in $\eta$ and 
plugging in the complicated expressions for the coefficients, we arrive to the following expression for the three-point function
with spatial boundary distance
\begin{equation} \label{T-spatiall-3point-perturbative}
S_{\rm total}  \, = \,S_{\rm KM} + \frac{\eta^4}{240 \, \Theta} \, \delta S^{(04)}
\end{equation}
where the expression for  $\delta S^{(04)}$ reads
\begin{eqnarray} \label{T-spatial-action_first_correction_general}
&&  \delta S^{(04)} \, =  \alpha_1^2 
\left(x_2 - x_3\right)^4 \left(\alpha_2 \, x_2^2 +\alpha_3 \, x_3^2\right) 
+ \, \alpha_2 \, \alpha_3 \left(x_2 - x_3\right)^2 \left(\alpha_3 \, x_2^4 +\alpha_2 \, x_3^4\right) + 
\alpha_1  \big(\alpha_3^2 \, x_2^4 \, x_3^2  
\nonumber \\[5pt]
&& +\alpha_2^2 \, x_3^4 \, x_2^2 \big) 
 + \, \alpha_1 \, \alpha_2 \, \alpha_3 \, \left( 2\, x_2^6 - 6 \, x_2^5 \, x_3 +  5 \, x_2^4 \, x_3^2 + 
5 \, x_3^4 \, x_2^2 - 6 \, x_3^5 \, x_2 + 2\, x_3^6\right) \, . 
\end{eqnarray}
The expression for $\Theta$ is the same as in  \eqref{3point-R-Theta}, under the interchange $\tau_i \rightarrow x_i$.
Moreover, the first order correction with spatial boundary distance in \eqref{T-spatial-action_first_correction_general} 
is identical with the first order correction with temporal boundary distance in
\eqref{T-temporal-action_first_correction_general},  when one interchanges $x_2$ with $\tau_2$ and $x_3$ with $\tau_3$. 
For $x_3 = 2 \, x_2$, it is possible to extend the perturbative calculation for the action to the eighth order in $\eta$. 
The result of the calculation reads
\begin{equation} \label{T-spatial-3point-perturbative_v2}
S_{\rm total}^{\, x_3\, = \, 2 \, x_2}  \, = \,S_{\rm KM}^{\, x_3\, = \, 2 \, x_2} + \frac{\eta^4\, x_2^4}{240\, \Theta} \,
 \delta S^{(04)}_{\, x_3\, = \, 2 \, x_2} -
 \frac{\eta^8\, x_2^8}{201600\, \Theta^5}  \, 
 \delta S^{(08)}_{\, x_3\, = \, 2 \, x_2} \, . 
\end{equation}
The coefficient  $ \delta S^{(04)}_{\, x_3\, = \, 2 \, x_2} $ is identical with $ \delta S^{(04)}_{\, \tau_3\, = \, 2 \, \tau_2} $ from  \eqref{T-temporal-3point-first-correction}, when one interchanges $x_2$ with $\tau_2$ and $x_3$ with $\tau_3$. Notice, however, that the next correction $\delta S^{(08)}_{\, x_3\, = \, 2 \, x_2}$ is different from the corresponding term $\delta S^{(08)}_{\, \tau_3\, = \, 2 \, \tau_2}$ in \eqref{T-temporal-3point-perturbative_v2} when one substitutes $\tau_{2,3}$ in the place of $x_{2,3}$. The expression for the coefficient $ \delta S^{(08)}_{\, x_3\, = \, 2 \, x_2} $ is listed in appendix 
\ref{appendix-3point-lengthy} in equation \eqref{T-spatial-3point-second-correction}. 

In the symmetric case  (i.e. $\Delta_1=\Delta_3=\Delta$) the perturbative result  \eqref{T-spatial-3point-second-correction}
becomes
\begin{eqnarray}
S_{\rm sym} &= & S_{\rm KM}^{\rm sym} + \frac{32 \Delta ^2-18 \Delta _2 \Delta +3 \Delta _2^2}{120 
\left(2 \Delta +\Delta_2\right)} \, \eta^4\, x_2^4 
\nonumber \\ 
& +&\frac{-26624 \Delta ^4+25512 \Delta _2 \Delta ^3-8196 \Delta _2^2 \Delta ^2+798 \Delta
   _2^3 \Delta +21 \Delta _2^4}{100800 \left(2 \Delta +\Delta _2\right){}^3}\, \eta^8\, x_2^8 \, .
\end{eqnarray}

\subsection{Finite temperature and R-charge chemical potential}
\label{3point-RT}

The most generic three-point correlator is when the background describes a CFT at both finite chemical potential and finite temperature. 
In this case the non-diagonal components of the metric \eqref{metric-general-v2} are non-zero and the calculation becomes 
much more involved. There are different scenarios that could be examined, depending on the type of operators that the 
geodesic approximation describes. In this section we will focus to the case of a three-point correlator with 1 returning and 
2 incoming geodesics. The returning and one of the incoming geodesics correspond to string solutions rotating in the 
internal space with the two angular momenta in the same direction (i.e. $\omega_2>0$ and  $\omega_3>0$ in the notation 
of section \ref{setup}), while for the second incoming geodesic the rotation in the internal space is along 
opposite directions (i.e. $\omega_2>0$ and  $\omega_3<0$). For the former case ($\omega_2>0$ and  $\omega_3>0$) the temporal boundary distance and the 
action are given by the expressions in equation \eqref{2point-tdot-positive-v1} while for the latter case ($\omega_2>0$ and  $\omega_3<0$) in equations 
\eqref{2point-tdot-negative-v1} and \eqref{2point-action-negative-v1}.

\begin{figure}[ht] 
   \centering
   \includegraphics[width=7.5cm]{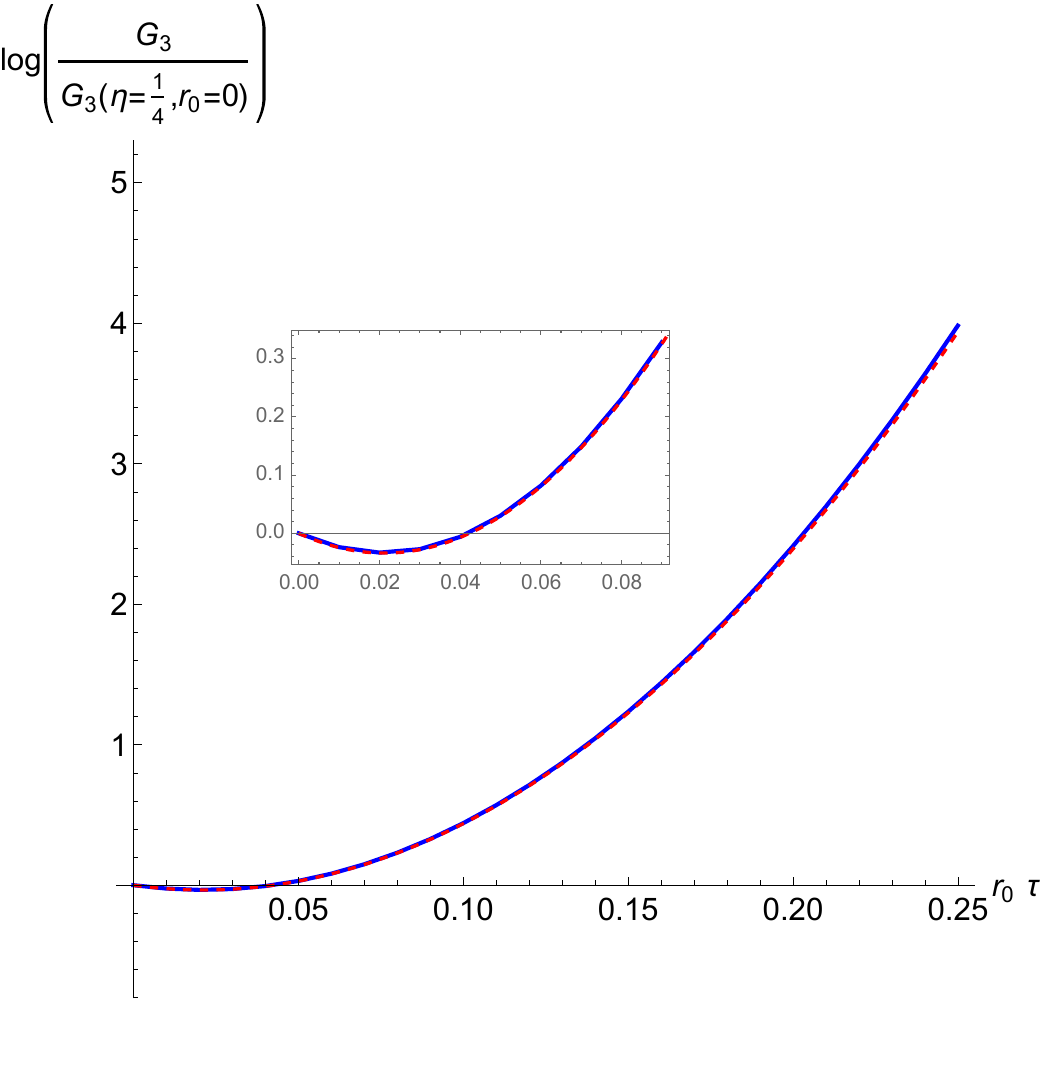}
   \includegraphics[width=7.5cm]{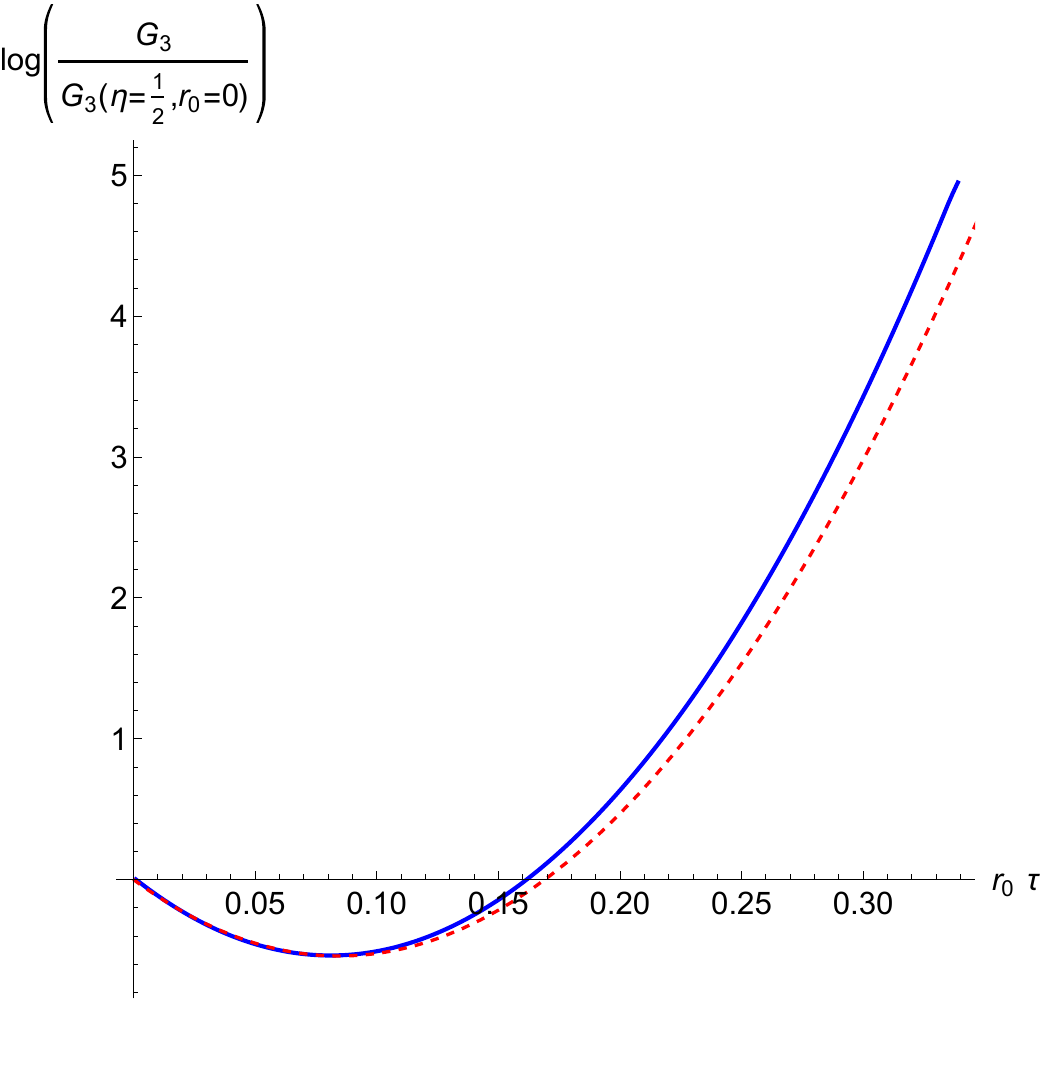}
    \caption{Plots of the logarithm of the three point function for arbitrary value of the parameter $r_0$ and fixed value of $\eta$ 
    ($\eta=1/4$ for the left part and $\eta=1/2$ for the right part) over the three point function for $r_0=0$ and $\eta=0$
    as a function of the dimensionless quantity $\tau T$. The operators are located at the points $\tau_1=0$, $\tau_2=1$ and 
    $\tau_3=2$ and as a result the parameter that changes in the horizontal axis is the value of $r_0$. The first operator 
    is returning with $\Delta_1=J_1= 155$, the second operator is incoming with $\Delta_2= 135$ and $J=2$ and the 
    third operator is also ingoing with $\Delta_3=J_3= 200$. We are not plotting the worldsheets of the three-point functions 
    since the configurations are qualitatively similar to those depicted in previous figures. 
    The blue curves are the result of the numerical computation, while for the red curves we have used the result of the 
    perturbative computation from equation \eqref{RT-3point-perturbative}. The perturbative result contains the first order 
    correction to the parameters $r_0$ and $\eta$ and the first mixed correction (i.e. $\eta^2 \,  r_0$) that is due to the presence 
    of the mixed term in the metric ansatz. To add more terms in the perturbative expansion one should restrict to the case 
    of the symmetric three-point functions that is discussed at the end of subsection \ref{3point-RT}.}
   \label{3point-RT-21}
\end{figure}

In the current setup, and in order to determine the meeting point for the three-point function, we have to define besides
the conformal dimensions of the three operators the difference of the two angular momenta for the operator that corresponds to rotation 
in the internal space along opposite directions. In figure \ref{3point-RT-21} we plot the three point function for arbitrary 
values of the rotation parameter $r_0$ and fixed value of the parameter $\eta$ ($\eta=1/4$ for the left part and 
$\eta=1/2$ for the right part) divided by the three point function for $r_0=0$ as a function of the dimensionless quantity 
$r_0 \tau$. Placing the operators at the boundary points $\tau_1=0$, $\tau_2=1$ and $\tau_3=2$, means that we are 
fixing $\tau=1$. This is analogous to the strategy we followed for the plot of the three-point functions in figures 
\ref{3point-R-21} and \ref{3point-R-12}. The blue curves are the result of the numerical computation, 
while for the red curves we have used the result of the perturbative computation that will be analyzed in the next subsection 
and is given by equation \eqref{RT-3point-perturbative}. The values of the conformal dimensions we have used are 
 $\Delta_1= 95$, $\Delta_2= 135$ and $\Delta_3= 200$, while the angular momentum is set to $J=2$. 
Similarly to the previous perturbative expressions, for small values of the rotation parameter $r_0$ 
the perturbative solution is close to the result of the numerical evaluation. The agreement between the numerical and the 
perturbative result is more pronounced in the left part of figure \ref{3point-RT-21}, where the value of $\eta$ is smaller. In 
any case, the perturbative result in \eqref{RT-3point-perturbative} is valid for small values of both $\eta$ and $r_0$.

A couple of important comments are in order:  To produce the plots of figure \ref{3point-RT-21} we have fixed the value 
of the parameter $\eta$ but not the value of the temperature, since the latter depends on both $r_0$ and $\eta$ 
as can be seen from equation \eqref{thermo-quantities}. To be precise, the temperature decreases as 
we move from the left to the right of each curve. To produce those plots we have used as a starting point for the 
numerical evaluation the position of the meeting point for finite $\eta$ and $r_0=0$ 
(from the analysis of the previous subsection) and then gradually  increase $r_0$. 
Keeping the temperature constant needs a more elaborate numerical analysis 
that is beyond the scope of this paper, but is a very interesting future direction. However, what we observe from both plots
in figure \ref{3point-RT-21} (in the left part we have zoomed for small values of $r_0$ to show that the qualitative behavior 
for different values of $\eta$ is similar) is that as the parameter $r_0$ increases 
(that is when the chemical potential increases and the temperature decreases) 
the value of the three-point function  decreased up to a point after which it started increasing. This a behavior that doesn't seem to 
depend on the value of the angular momentum (we have checked for different values of $J$) and is supported by 
the perturbative computation of equation \eqref{RT-3point-perturbative}. 
Using the perturbative computation and trading $\eta$ with the temperature we can check that the behavior we see in 
figure \ref{3point-RT-21} remains intact when the temperature is fixed and takes small values.  
A numerical computation could address the 
same question when the temperature increases, since in this case the perturbative computation is not valid any more. The approximate value of the minimum of the action can be easily computed from \eqref{RT-3point-perturbative} to be at
\begin{equation} 
r_0^{(min)}=-\frac{\eta^2}{2 \, \Theta^3} \, \frac{\delta S^{(12)}}{\delta S^{(20)} }\, .
\end{equation}
It would be interesting to understand better the physical significance of the different behavior of the three-point function 
in the presence of both chemical potential and temperature compared to the cases where only temperature or 
only chemical potential are present.\\

\paragraph{Perturbative computation}

\noindent

\noindent
To complement the numerical computation we have performed and discussed previously, we perform the perturbative 
computation of three-point function around the $AdS_5$ result \cite{Klose:2011rm,Minahan:2012fh} (i.e. $\eta=r_0=0$).
To this end, we consider three operators with conformal dimensions and angular momenta as follows: 
$\Delta_1=J_1$, ($\Delta_2$, $J_2=J$) and $\Delta_3=J_3$, that are placed at the boundary points $(\tau, z)= (0, 0)$, 
$(\tau_2, 0)$ and $(\tau_3, 0)$ respectively. We consider the case in which the first operator is returning 
and the other two operators incoming. \footnote{Let us remind the reader that $J_2=J$ is the difference between the two angular momenta of the operator corresponding to \eqref{Delta-J-negative}.}

To solve the equations that determine the meeting point, we expand around the zero R-charge and zero temperature  
solution in the following way 
\begin{eqnarray} \label{RT-3point-expansion-ansatz}
&& 
\tau_{I} = \tau_0 +  r_0^2 \, \delta^{(20)} \tau + r_0 \, \eta^2  \, \delta^{(12)} \tau + \eta^4 \, \delta^{(04)} \tau  \, ,  \quad 
z_{I} = z_0+  r_0^2 \, \delta^{(20)} z + r_0 \, \eta^2  \, \delta^{(12)} z + \eta^4 \, \delta^{(04)} z
 \nonumber \\[5pt]
 &&
 \qquad  \qquad \& \quad  \mu_i \, = \, \mu_{i  0} \,  + \, r_0^2 \, \delta^{(20)} \mu_i + r_0 \, \eta^2  \, \delta^{(12)} \mu_i
 + \, \eta^4 \, \delta^{(04)} \mu_i  
 \quad i =1,2,3 \, .
\end{eqnarray}
Notice that the first number inside the parenthesis (i.e. in the index) depicts the power of the parameter $r_0$
and the second the power of the parameter $\eta$. Solving perturbatively the system of equations, the terms with indices 
(20) and (04) end up to be the same with those we have computed in previous perturbative expansions (in the absence of 
R-charge or temperature). For this reason we use the same symbols. The new ingredient in the computation is coming 
from the terms with index (12). Even if we have analytic expressions for those terms, they are very long and non-illuminating 
to present them.

The total action that will be exponentiated comprises of one returning and two ingoing pieces (as in 
\eqref{R-action-perturbative_v0}) and substituting the expansion ansatz of \eqref{RT-3point-expansion-ansatz} 
we arrive to the following expression for the three-point function
\begin{equation} \label{RT-3point-perturbative}
S^{\rm RT}_{\rm total}  \, = \,S_{\rm KM}  - \frac{r_0^2}{12} \, \delta S^{(20)} - 
\frac{\eta^4}{80 \, \Theta} \, \delta S^{(04)} - \frac{r_0 \, \eta^2}{12 \, \Theta^3}  \, \delta S^{(12)} \, . 
\end{equation}
The expressions for the coefficients $\delta S^{(20)}$ and $\delta S^{(04)}$ have been already calculated during the perturbative expansions for zero temperature and zero R-charge respectively, while the 
expression for the coefficient $\delta S^{(12)}$ is listed in appendix 
\ref{appendix-3point-lengthy} in equation \eqref{RT-temporal-3point-mixed}.
Notice that in the symmetric case (i.e. $\alpha _3=\alpha_1$ and $\tau_3 = 2 \, \tau_2 $) the coefficient $\delta S^{(12)}$ vanishes. 


\paragraph{Perturbative computation of the symmetric three-point function}

\noindent

\noindent
Trying to calculate more terms in the perturbative expansion of the three-point function when both temperature and 
chemical potential are present becomes tedious. Consequently, we turn to the case of the symmetric configuration. In this case the operators 1 and 3 are 
returning with $\omega_3>0$ for each one of them and the operator 2 is incoming with $\omega_3<0$. 
The expansion ansatz in this case is
\begin{eqnarray} \label{RT-3point-expansion-symmetric-ansatz}
&& 
\tau_{I} = \tau_0 + r_0 \, \eta^2  \, \delta^{(12)} \tau  + r_0^3 \, \eta^2  \, \delta^{(32)}\tau\, ,  \quad 
\mu_2 = r_0 \, \eta^2  \, \delta^{(12)} \mu_2 + r_0^3 \, \eta^2  \, \delta^{(32)} \mu_2  
 \nonumber \\[5pt]
 &&
z_{I} = z_0+  r_0^2 \, \delta^{(20)} z+ r_0^4 \, \delta^{(40)} z + \left[ \delta^{(04)} z+r_0^2 \, \delta^{(24)} z \right]\eta^4  
\\[5pt]
 &&
 \mu_1 =  \mu_{1  0} +  r_0^2 \, \delta^{(20)}  \mu_1 + r_0^4 \, \delta^{(40)}  \mu_1  + 
 \left[ \delta^{(04)}  \mu_1 + r_0^2 \, \delta^{(24)}  \mu_1  \right]\eta^4 
+  r_0 \, \eta^2  \, \delta^{(12)}  \mu_1  + r_0^3 \, \eta^2  \, \delta^{(32)}  \mu_1  
\nonumber \\[5pt]
 &&
 \mu_3 =  - \mu_{1  0} -  r_0^2 \, \delta^{(20)}  \mu_1 - r_0^4 \, \delta^{(40)}  \mu_1  - 
 \left[ \delta^{(04)}  \mu_1 + r_0^2 \, \delta^{(24)}  \mu_1  \right]\eta^4 
+  r_0 \, \eta^2  \, \delta^{(12)}  \mu_1  + r_0^3 \, \eta^2  \, \delta^{(32)}  \mu_1  \,. 
\nonumber
\end{eqnarray}
An important comment is in order: If one looks in the expansion ansatz of  
\eqref{RT-3point-expansion-symmetric-ansatz} it seems that some terms are missing. For example, in the expansion of
$z_I$ the mixing terms $ \delta^{(12)} z$ and $ \delta^{(32)} z$ are not present. The reason for their absence is that 
from the perturbative solution of the system of equations that determine the meeting point, their value came out to 
be zero. The expressions for the remaining (non-zero) coefficients are listed in appendix 
\ref{appendix-symmetric-expansion}. 
The total action that will be exponentiated will have two returning components and one incoming component
\begin{equation} \label{RT-3point-perturbative_symmetric-def}
S_{\rm total}^{\rm sym}  = S_{\rm ret} (\mu_1, z_I) + S_{\rm in} (\mu_2, z_I) +S_{\rm ret} (\mu_3, z_I) \, . 
\end{equation}
Expanding \eqref{RT-3point-perturbative_symmetric-def} we obtain the following expression for the symmetric 
three-point function
\begin{eqnarray} \label{RT-3point-perturbative_symmetric}
&& S_{\rm total}^{\rm sym}  \, = \,S_{\rm KM}^{\rm sym} -
\frac{1}{40} \, \frac{2 \, \alpha _1^2+7 \, \alpha _1 \, \alpha _2+8 \, \alpha _2^2}{2 \, \alpha _1+\alpha _2} \, \eta^4 \,  \tau _2^4
- \frac{1}{6} \left(\alpha _1+ 2\, \alpha _2 \right)r_0^2 \,  \tau _2^2
\nonumber \\[5pt]
&&
\qquad -\Bigg[\frac{132 \, \alpha _1^4+716 \, \alpha _2 \, \alpha _1^3+1737 \, \alpha _2^2 \, \alpha _1^2+1896 \, 
\alpha _2^3 \, \alpha _1+840 \, \alpha _2^4}{1680 \, \alpha _1 
\left(2 \, \alpha_1+\alpha _2\right)^2}
\nonumber \\[5pt]
&&
\qquad + \frac{ \left(J-\alpha _1\right) \alpha _2^2 \Big[24 \, \alpha _1^3+24 \, \alpha _2^3+8 \, \alpha _1 \, \alpha _2 
\left(J+10 \,\alpha _1\right)+\alpha _2^2 \left(7\, J +79 \, \alpha _1\right)\Big]}{48 \, \alpha _1 
\left(2 \, \alpha _1+\alpha _2\right)^4}\Bigg] \, r_0^2 \,  \eta^4 \, \tau _2^6
\nonumber \\[5pt]
&&
\qquad -
\frac{1}{180} \, \frac{4 \, \alpha _1^2+19 \, \alpha _1 \, \alpha _2+16 \, 
\alpha _2^2}{2 \, \alpha _1+\alpha _2} \, r_0^4 \,  \tau _2^4 \, . 
\end{eqnarray}
Setting $\alpha_1=\alpha_3=\Delta_2$ and $\alpha_2=2\, \Delta - \Delta_2$ it is possible to express 
\eqref{RT-3point-perturbative_symmetric} entirely in terms of the conformal dimensions.
Finally, notice that using \eqref{thermo-quantities} it is straightforward to express the result for the two-point function in terms of the field theory quantities $T$ and $\Omega$.


\subsection{Contributions from the internal space}
\label{contribution-sphere}

In this section we will discuss the validity of our approximation, as well as additional contributions coming from the internal space. 
To start, notice that the complete three-point correlator at zero temperature and chemical potential is not given by the first term of \eqref{RT-3point-perturbative} but receives additional contributions from the internal space, i.e. the $S^5$ \cite{Buchbinder:2011jr}. The method initiated in \cite{Georgiou:2022ekc} and further developed in the present paper, while it takes into account the motion of the point-like string in the internal space, it does not fully capture the complete three point function. This is obvious since, as mentioned above, our zero-th order result captures only the $AdS_5$ part of the correlator.

To see what happens in some more detail let us focus on the leading term of the expansion \eqref{RT-3point-perturbative} that depends on both deformations, i.e. the term proportional to $r_0 \, \eta^2$. Since our calculation is set up as perturbation around the $AdS_5\times S^5$ result we expand the metric around this and keep only the terms proportional to $r_0 \, \eta^2$. To calculate the Laplacian, we expand the determinant of the metric (that does not receive corrections up to the aforementioned order)
\begin{equation}
 \sqrt{-G} =  \sqrt{-g_{AdS}} \sqrt{g_{S^5}} + {\cal O} \left(r_0^2 \right)
\end{equation}
while the expansion of the inverse metric reads
\begin{equation}
g^{\mu \, \nu} =  g^{\mu \, \nu}_{AdS \times S^5} + g^{t\, \phi_2} +g^{t\, \phi_3} \quad {\rm with} \quad 
g^{t\, \phi_2} = g^{t\, \phi_3} = - r_0 \, \eta^2 \, z^4 \, . 
\end{equation}
Substituting all those expansions, 
the Laplacian for a mass-less scalar field then becomes
\begin{equation}\label{Lapl}
    \nabla^2_{AdS} \Phi + \nabla^2_{S^5} \Phi = 2 \, r_0 \, \eta^2\, z^4
    \Big(\partial_t \partial_{\phi_2} \Phi+ \partial_t \partial_{\phi_3} \Phi\Big)
\end{equation}
where
\begin{equation}
    \nabla^2_{AdS} \Phi = \frac{1}{\sqrt{-g_{AdS}}} \partial_{\mu} 
    \Big[\sqrt{-g_{AdS}} \, g^{\mu \nu} \, \partial_{\nu} \Phi\Big]
    \quad \& \quad 
    \nabla^2_{S^5} \Phi = \frac{1}{\sqrt{g_{S^5}}} \partial_{\alpha} 
    \Big[\sqrt{g_{S^5}} \, g^{\alpha \beta} \, \partial_{\beta} \Phi\Big].
\end{equation}
As usual the mass of the particle will come from the $S^5$ part that is the second term in the left hand side of \eqref{Lapl}.

Since the right hand side of \eqref{Lapl} does not depend explicitly on the angles of the $S^5$ one can make the following ansatz
\begin{equation}\label{correction}
    \Phi =  \Phi_0 (x^{\mu}) \, \Upsilon_{\Delta}(\Omega) + r_0 \, \eta^2 \, \Phi_1 (x^{\mu}, \Omega)
     \quad {\rm with} \quad 
     \Phi_1 (x^{\mu}, \Omega) = \Phi_{1,\Delta}^{(1)} (x^{\mu})\, \Upsilon_{\Delta}(\Omega),
\end{equation}
where $\Upsilon_{\Delta}(\Omega)$ is the spherical harmonic corresponding to the operator we are using.
After using the fact that the wave-function $\Phi_0 (x^{\mu}) \, \Upsilon_{\Delta}(\Omega)$ satisfies, by assumption, the zero-th order differential equation we obtain for the correction $\Phi_{1,\Delta}^{(1)} (x^{\mu})$ an equation that does not depend at all on the coordinates of the internal space.

\begin{equation}
   \frac{1}{ \Phi_{1,\Delta}^{(1)} \, \Upsilon_{\Delta} } \Bigg[\nabla^2_{AdS} (\Phi_{1,\Delta}^{(1)}\, \Upsilon_{\Delta}) + \nabla^2_{S^5}  (\Phi_{1,\Delta}^{(1)}\, \Upsilon_{\Delta})\Bigg]  = 
   \frac{2\, z^2 }{ \Phi_{1,\Delta}^{(1)} \, \Upsilon_{\Delta} } 
    \Bigg[\partial_t \Phi_{1,\Delta}^{(1)}  \partial_{\phi_2} \Upsilon_{\Delta}+ \partial_t \Phi_{1,\Delta}^{(1)}  \partial_{\phi_3} \Upsilon_{\Delta}\Bigg]
\end{equation}
or
\begin{equation}\label{nabla}
 \nabla^2_{AdS} \Phi_{1,\Delta}^{(1)} + 
  \Delta (\Delta-4)\Phi_{1,\Delta}^{(1)}= 
   2\, i\, z^2 \, \partial_t \Phi_{1,\Delta}^{(1)}  \left(n_2 + n_3\right), 
\end{equation}
where we have used that $ \Upsilon_{\Delta}\sim e^{i (n_2 \phi_2+n_3 \phi_3)}$ 
and that $\nabla^2_{S^5}  \Upsilon_{\Delta}=\Delta (\Delta-4)  \Upsilon_{\Delta}$.
The important point in \eqref{nabla} is that it is a function of ($t,z,x_1,x_2,x_3$) only. As a result, and up to order $r_0 \, \eta^2$, the result for the overlap in the sphere is the same as in the un-deformed case giving the following contribution, see \eqref{sphere_contribution} below. 
In conclusion, the sphere contribution is given by the logarithm of \eqref{sphere_contribution} and it should be added to the perturbative computation of \eqref{RT-3point-perturbative}.

In fact, one may argue that the same result is valid as long as we restrict ourselves to small values of the temperature and the chemical potential and to operators of large dimensions and angular momenta.
The argument goes as follows. When considering the next terms in the expansion in terms of $r_0$ and $\eta$ the equation for the scalar becomes 
\begin{eqnarray}\label{Lapl2}
    \nabla^2_{AdS} \Phi + \nabla^2_{S^5} \Phi &=& 2 \, r_0 \, \eta^2\, z^4
    \Big(\partial_t \partial_{\phi_2} \Phi+ \partial_t \partial_{\phi_3} \Phi\Big) + \eta^4 \, z^6 \, \partial^2_t \Phi - r_0^2 \, z^2 \, 
    \partial^2_{\phi_1} \Phi \nonumber \\
    && +\, 2 \, r_0^2 \, \eta^4 \, z^6\, \partial_{\phi_2} \partial_{\phi_3}\Phi + \left(r_0^2 \, \eta^4 \, z^6 + \frac{r_0^2\, z^2}{\sin^2 \psi}\right) \partial^2_{\phi_2} \Phi \nonumber \\
    && + \left(r_0^2 \, \eta^4 \, z^6 + \frac{r_0^2\, z^2}{\cos^2 \psi}\right) \partial^2_{\phi_3} \Phi+ \frac{r_0^2\, z^2}{\sin 2\psi} \partial_{\psi} \Big(\sin 2\psi \, \partial_{\psi} \Phi\Big) \nonumber \\
    && +z^5 \partial_z \Bigg[\Big(z\, \eta^4 + \frac{r_0^2}{z} - r_0^2 \, \eta^4 \, z^3\Big)\partial_z\Phi\Bigg] \, . 
\end{eqnarray}
Notice that the right-hand side of \eqref{Lapl2} now depends on the angles of the 5-sphere. 
However, in the limit of large dimensions, in which the operator is approximated by a classical trajectory, one can proceed as follows: First we multiply equation \eqref{Lapl2} by $\Upsilon_{\Delta'}(\Omega)$ and integrate over the the volume of $S^5$, namely $d\Omega=\cos^3{\theta}\sin{\theta}\sin{\psi}\cos{\psi}\, d \theta d \psi d \phi_2 d \phi_3 d \phi_1$.
At this point, let us make the identification of the spherical harmonics that approximates the classical solutions \eqref{ABconstants-positive} and \eqref{ABconstants-negative} which are dual to the operators in \eqref{ops}. These are of the form \cite{Hernandez:2005xd}
\begin{equation}\label{harmonic}
    \Upsilon_{\Delta}(\Omega)\sim e^{i n_2 \phi_2+i n_3 \phi_3} \cos^{l}\theta \,\sin^{|n_2|}\psi \,\cos^{|n_3|}\psi,
\end{equation}
 where $l=|n_2|+|n_3|$.
To see this notice that, for large values $n_2$ and $n_3$ and upon identification of $\frac{n_2}{n_3}$ with  $\frac{\omega_2}{\omega_3}$, the above spherical harmonic is a narrow function with its peak (maximum) situated at precisely the values of the classical solutions, namely at $\theta=0$ and $\psi=\frac{1}{2}\arccos \psi_0$, where $\psi_0$ is given by \eqref{ABconstants-positive} and \eqref{ABconstants-negative}. \footnote{For the angle theta this statement is obvious. For the angle $\psi$ one can easily verify that the derivative of the spherical harmonic with respect to $\psi$ becomes zero at the point $\psi=\frac{1}{2}\arccos \psi_0$. }
After acting with the sphere derivatives all the integrals in \eqref{Lapl2} will be of the form
\begin{equation}\label{integral}
   I= \int d\Omega\,\Upsilon_{\Delta'}(\Omega) f(\theta,\psi)\Upsilon_{\Delta}(\Omega),
\end{equation}
with $\Upsilon_{\Delta}(\Omega)$ and   $\Upsilon_{\Delta'}(\Omega)$ being of the form \eqref{harmonic}. The integral \eqref{integral} can now be evaluated using the saddle point approximation since $n_2$ and  $n_3$ are big integers. 
The integrals over $\phi_2$ and  $\phi_3$ will impose the relations $n'_2=n_2$ and  $n'_3=n_3$. As a result the integral becomes
\begin{equation}\label{integral-saddle}
I=f\left(\theta=0,\psi=\frac{1}{2}\arccos \psi_0\right)\Bigg({|n_2|\over|n_2|+|n_3|}\Bigg)^{n_2}\Bigg({|n_3|\over|n_2|+|n_3|}\Bigg)^{n_3}\, .
\end{equation}
Consequently, the dependence on the sphere coordinates will drop out and \eqref{Lapl2} will end up to be an equation for ($t,x_1,x_2,x_3,z$), which depends on the sphere coordinates only through the constant values of the angles at which the classical solutions are situated. This equation can now be solved perturbatively in powers of $r_0$ and $\eta$ using an ansatz like the following
\begin{eqnarray}\label{correction2}
   \Phi &=&  \Phi_0 (x^{\mu}) \, \Upsilon_{\Delta}(\Omega) + \Bigg(r_0 \, \eta^2 \, \Phi_1^{(1,2)} (x^{\mu})+\eta^4 \, \Phi_1^{(0,4)} (x^{\mu})+r_0^2 \, \Phi_1^{(2,0)} (x^{\mu})
   \nonumber \\
   && +r_0^2 \, \eta^4 \, \Phi_1^{(2,4)} (x^{\mu})+\cdots\Bigg)\, \Upsilon_{\Delta}(\Omega) \, . 
\end{eqnarray}
To summarise, the solution of the resulting differential equation will not depend on the coordinates of the sphere and in the large-$\Delta$ limit will be approximated by the three joining geodesics (extrema) of \eqref{NG}. 
This is the essence of our method.
Finally, as long as the parameters $r_0$ and $\eta$ are treated as small perturbations the overlap over the sphere $\int d\Omega\,\Upsilon_{\Delta_1}(\Omega) \Upsilon_{\Delta_2}(\Omega)\Upsilon_{\Delta_3}(\Omega)$ will be calculated by saddle point approximation and will be 
\begin{eqnarray} \label{sphere_contribution}
&&\int d\Omega\,\Upsilon_{\Delta_1}(\Omega) \Upsilon_{\Delta_2}(\Omega)\Upsilon_{\Delta_3}(\Omega) = \nonumber \\
&&\Bigg({N_2\over N_2+ N_3 }\Bigg)^{N_2}\Bigg({ N_3 \over N_2 + N_3 }\Bigg)^{N_3}\delta_{n_2^{(1)}+ n_2^{(2)}+ n_2^{(3)},0}\delta_{n_3^{(1)}+ n_3^{(2)}+ n_3^{(3)},0}
\end{eqnarray}
where
\begin{equation} 
N_2 = |n_2^{(1)}|+ |n_2^{(2)}|+ |n_2^{(3)}| \quad \& \quad 
N_3 = |n_3^{(1)}|+ |n_3^{(2)}|+ |n_3^{(3)}|
\end{equation}
and the relation connecting the anomalous dimensions with the spin is 
\begin{equation}
    \Delta_i = |n_2^{(i)}|+ |n_3^{(i)}| + 4 \quad {\rm for} \quad i=1,2,3 \, .
\end{equation}
In conclusion, the logarithm of \eqref{sphere_contribution} should be added to the perturbative computation of \eqref{RT-3point-perturbative}.


\section{Conclusions}

In this paper we continue the study of conformal field theories in the presence of finite temperature and density, 
that was initiated in \cite{Georgiou:2022ekc}. The dual string theory background, that is presented in 
section \ref{intro}, is generated by non-extremal and rotating D3-branes with two equal angular momenta. 
The main focus of the analysis is on the holographic calculation of three point correlation functions of operators 
with large conformal dimensions. In order to complete such a task 
we generalised the motion of the particle in the internal space that we considered in  \cite{Georgiou:2022ekc}.
More specifically the motion of the point like string is now along two of the angles of the deformed 5-sphere. This 
generalization is imposed by the conservation of the R-charge density. For the class of solutions we are using, when only one isometry is turned on, the 
corresponding three-point correlation function becomes extremal. In the strong coupling regime it factorizes 
into the product of two two-point functions. To go beyond extremality, one needs to consider strings with two angular momenta.
Having introduced a new class of classical string solutions, in section \ref{setup} we identify the dual operators and we extend the general method for the 
holographic calculation of correlation functions that was introduced in \cite{Georgiou:2022ekc}. The novel feature, that this 
new class of solutions is highlighting, is that the on-shell action (the exponential of which is related to the value 
of the correlation function) depends not only on the conformal dimension of the operator but 
also on the difference of the two angular momenta. This is something that, to the best of our knowledge, appears for the 
first time in the literature so far and is due to the fact the gravity background distinguishes the direction of rotation, i.e.  it is 
not invariant under $\varphi \rightarrow - \varphi$. In the second part of section \ref{setup}, we revisit the general method of 
calculating three-point functions of operators with large conformal dimensions, where the bulk-to-boundary propagator is 
approximated by the exponential of minus the mass times the geodesic length. This geodesic corresponds to the trajectory 
of a particle traveling between the boundary point and the bulk interaction point.

To fully understand the significance of this extra parameter (i.e. difference of the angular momenta), in section 
\ref{2point-functions} we present the holographic calculation of two-point correlation functions. When the rotation around the 
two isometries of the deformed 5-sphere is along different directions, the result for the two-point correlator depends explicitly on the difference of 
the angular momenta (see equation  \eqref{2point-action-negative_v2}). In the case of rotation along the same direction, 
one re-obtains the result of  \cite{Georgiou:2022ekc}. It should be emphasized here that in order to observe the two different
behaviors of the two-point correlator, both the temperature and the chemical potential should be turned on. 
The reason is that the non-diagonal terms in the metric are absent if either the temperature or the chemical potential 
are switched off.  It is precisely the existence of these terms which causes the two different behaviors to appear.  

 The main results of the current paper are presented in section \ref{3point-functions}. In this section we split the analysis 
in three cases: In the first two subsections we present the holographic calculation for the three-point correlators, 
when the background contains either only R-charge density or only temperature, but not both while in the last subsection we 
present the most general case when the background is in full generality and contains both finite chemical potential
and finite temperature. In subsection \ref{contribution-sphere}, we analyze the contribution of the internal space to the three-point correlator.

To determine the bulk interaction point we solve an algebraic system of equations, 
both numerically and perturbatively around the undeformed background, for all three cases. 
Here we summarize the most important  findings: 
The numerical computation stops giving real solutions for the coordinate of the interaction point 
when there is considerable distance between the world-sheet of the three-point point function and the singularity/horizon. 
This suggests that the triangle inequality of \cite{Klose:2011rm} for the conformal dimensions of the operators
needs to be modified or that another saddle point dominates after the critical point at which the solution becomes complex. It would be interesting to clarify which of the two is happening.

In all the cases we have considered the perturbative, analytic solution is very close to the numerical result, even when the small parameter that we expand 
($r_0$ or $\eta$) is not so small. This can be seen from all the plots that we present in section \ref{3point-functions}.
Furthermore, the expression for the analytic perturbative three-point function is the same regardless of the shape of the world-sheet (one returning and two in going or two returning and one in going geodesics). 
Another important observation from the plots of the three point function (when the background 
contains either temperature or density but not both) is that it is always a monotonic function. 
In particular, for the case with temperature only, the three-point function is monotonically increasing for a temporal  distance between the boundary points while it is monotonically decreasing 
when the boundary distance in spatial. 
However, when both temperature and chemical potential are present, an unexpected behavior arises. The 
logarithm of the three point function is no longer a monotonic function. For fixed finite temperature and small values of the
chemical potential, it starts decreasing and after some point the behavior changes and becomes increasing, i.e. there is a 
minimum for the value of the three-point function. Numerical and perturbative computation are in agreement as far as the 
existence of this minimum, for all the values of the conformal dimensions or of the angular momentum that we have considered. It 
would be interesting to investigate more on this different behavior when both temperature and chemical potential are present.
Given that the aforementioned behaviors have been observed for a certain range of the conformal dimensions and of the positions of the operators, it would be interesting to scan the full parametric space to investigate if these behaviors persist. 
Finally, it would be very interesting to investigate if the features of the correlators in the presence of temperature and/or chemical potential that were mentioned above survive at the weak coupling regime. To this end perturbative calculations of the three-point functions will be required.


\section*{Acknowledgments}

The research work of this project was supported by the Hellenic Foundation
for Research and Innovation (H.F.R.I.) under the “First Call for
H.F.R.I. Research
Projects to support Faculty members and Researchers and the procurement of
high-cost research equipment grant” (MIS 1857, Project Number: 16519).


\appendix


\section{Lengthy three-point coefficients}
\label{appendix-3point-lengthy}

In this appendix we list several analytic, but lengthy and not particularly illuminating,  three-point coefficients, 
for the interested reader that would like to reproduce those computations. 

\begin{eqnarray}  \label{T-spatial-3point-second-correction}
&&\delta S^{(08)}_{\, x_3\,  = \, 2 \, x_2}= 13 \left(\alpha _2+4 \, \alpha _3\right)^5 \alpha _1^6+2 
\Big(1664 \, \alpha _2^6+24519 \, \alpha _3 \, \alpha _2^5+115850 \, \alpha _3^2 \, \alpha _2^4+
217920 \, \alpha_3^3 \, \alpha _2^3 
\nonumber \\[5pt]
&&
+ 205760 \, \alpha _3^4 \, \alpha _2^2+97536 \, \alpha _3^5 \, \alpha _2+6656 \, \alpha _3^6\Big) \alpha _1^5+ 
5 \, \alpha _2 \, \alpha _3 \Big(3328 \, \alpha _2^5+32903 \, \alpha _3 \, \alpha _2^4+100744 \, \alpha _3^2 \, 
\alpha _2^3 
\nonumber \\[5pt]
&&
+138688 \, \alpha _3^3 \, \alpha _2^2+82304 \, \alpha _3^4 \, \alpha _2+3328 \, \alpha_3^5\Big) \alpha _1^4+
20 \, \alpha _2^2 \, \alpha _3^2 \Big(1664 \, \alpha _2^4+11037 \, \alpha _3 \, \alpha _2^3+25186 \, \alpha _3^2 
\, \alpha _2^2
\nonumber \\[5pt]
&&
+21792 \, \alpha_3^3 \, \alpha _2+416 \, \alpha _3^4\Big) \alpha _1^3+5 \, \alpha _2^3 \, \alpha _3^3 
\Big(6656 \, \alpha _2^3+32903 \, \alpha _3 \, \alpha _2^2+46340 \, \alpha _3^2 \, \alpha _2+416 \, \alpha _3^3\Big) 
\alpha _1^2 
\nonumber \\[5pt]
&&
+2 \, \alpha _2^4 \, \alpha _3^4 \left(8320 \, \alpha _2^2+24519 \, \alpha _3 \, \alpha _2+130 \, \alpha _3^2\right)
   \alpha _1+13 \, \alpha _2^5 \, \alpha _3^5 \left(256 \, \alpha _2+\alpha _3\right) 
\end{eqnarray}

\begin{eqnarray}   \label{RT-temporal-3point-mixed}
&& \delta S^{(12)} = \alpha _2^3 \, \alpha _3^3 \left(\tau _2-\tau _3\right)^6 
\left(\alpha _3 \, \tau _2^3+\alpha _2 \, \tau _3^3\right) +\alpha _1 \, \alpha _2^2 \, \alpha _3^2 
\left(\tau_2-\tau _3\right)^4 
\Bigg[3 \, \alpha _3^2 \, \tau _3^2 \, \tau _2^3+3 \, \alpha _2^2 \, \tau _3^3 \, \tau _2^2 
\nonumber\\
&&
+ \alpha _2 \, \alpha _3 \left(\tau _2+\tau _3\right) 
\left(\tau_2^2-\tau _3 \, \tau _2+\tau _3^2\right) 
\left(4 \, \tau _2^2-5 \, \tau _3 \, \tau _2+4 \, \tau _3^2\right)\Bigg] 
+\alpha _1^4 \left(\tau _2-\tau _3\right)^3
 \left(\alpha _2 \, \tau _2^2-\alpha _3 \, \tau _3^2\right) \times
\nonumber \\
 && 
 \left(\alpha _2^2 \, \tau _2^4+4 \, \alpha _2 \, \alpha _3 \,  \tau _3^2 \, \tau _2^2+\alpha _3^2 \, \tau _3^4 \right) -
 3 \, \alpha _1^2 \, \alpha _2 \, \alpha _3 \left(\tau _2-\tau _3\right)^2 
\Bigg[-2 \, \alpha _2^2 \, \alpha _3 \,  \tau _2^7+5 \, \alpha _2^2 \,  \alpha _3 \, \tau _3  \, \tau _2^6
\nonumber\\[5pt]
&&
-3 \, \alpha _2 \, \alpha _3 \left(2 \, \alpha _2+\alpha _3\right) \tau _3^2\,  \tau _2^5+\alpha _2 
\left(\alpha _2^2+4 \, \alpha _3 \, \alpha _2+ 5 \, \alpha _3^2\right) \tau _3^3 \, \tau _2^4+
\alpha _3 \left(5 \, \alpha _2^2+4 \,  \alpha _3 \, \alpha _2+\alpha _3^2\right) \tau _3^4 \,  \tau _2^3
\nonumber\\[5pt]
&&
-3 \, \alpha _2 \, \alpha _3 \left(\alpha _2+2 \, \alpha_3\right) \tau _3^5 \, \tau _2^2+
5 \, \alpha _2 \, \alpha _3^2 \, \tau _3^6 \, \tau _2-2 \, \alpha _2 \, \alpha _3^2 \, \tau _3^7\Bigg]
-\alpha _1^3 \Bigg[\alpha _2^4 \,  \tau _3^3 \, \tau _2^6-\alpha _2^3 \, \alpha _3 
\left(\tau _2-2 \tau _3\right)  \times
\nonumber\\
&&
\left(\tau _2^2-\tau _3 \, \tau _2+\tau _3^2\right) \left(4 \, \tau _2^2-3 \, \tau _3 \, \tau _2+3 \, \tau_3^2\right) \tau _2^4
+\alpha _3^4 \, \tau _3^6 \, \tau _2^3-3 \, \alpha _2^2 \, \alpha _3^2 \, \tau _3^2 
\left(\tau _2+\tau _3\right) \Bigg(3 \, \tau _2^4-13 \, \tau _3 \, \tau_2^3
\nonumber \\
&&
+19 \, \tau _3^2 \, \tau _2^2-13 \,  \tau _3^3 \, \tau _2+3 \, \tau _3^4\Bigg) \tau _2^2 
+\alpha _2 \, \alpha _3^3 \left(2 \, \tau _2-\tau _3\right) \tau _3^4 
\left(\tau_2^2-\tau _3 \, \tau _2+\tau _3^2\right) \left(3 \, \tau _2^2-3 \, \tau _3 \, \tau _2+4 \, \tau _3^2\right)\Bigg]
\nonumber\\
&&
- \, \alpha _2^2 \, \tau _2^3 \left(\tau _2-\tau _3\right){}^3 
\Big[\left(\alpha _1+\alpha _3\right) \tau _2-\alpha _3 \tau _3\Big] 
\Bigg[\alpha_2 \left(\alpha _1+\alpha _3\right){}^2 \tau _2^2- 2 \, \alpha _2 \, \alpha _3 
\left(\alpha _1+\alpha _3\right) \tau _3 \, \tau _2
\nonumber\\
&&
+ \, \alpha _3 \Big[3 \, \alpha _1 \left(\alpha _1+\alpha _2\right)+ \left(3 \, \alpha _1+\alpha _2\right) \alpha _3\Big] 
\tau _3^2\Bigg]  \left(2 \, J -\alpha _1-\alpha_3\right)
\end{eqnarray}


\section{Perturbative coefficients of the symmetric three-point function}
\label{appendix-symmetric-expansion}

In this appendix we list the coefficients of the perturbative expansion 
\eqref{RT-3point-expansion-symmetric-ansatz}. These are $\delta^{(20)} z$ and $\delta^{(20)}  \mu_1$ with expressions
\begin{equation}
\delta^{(20)} z = \frac{ \tau_2^3  }{6} \sqrt{\frac{\alpha_2}{2\, \alpha_1 + \alpha_2}}\, , \quad 
\delta^{(20)}  \mu_1 = - \frac{\tau_2}{3} \, \frac{\alpha_1+ 2\, \alpha_2}{\alpha_1+ \alpha_2}
\end{equation}
$\delta^{(40)} z$ and $\delta^{(40)}  \mu_1$ with expressions
\begin{equation}
\frac{\delta^{(40)} z}{\delta^{(20)} z} = \frac{ \tau_2^2  }{20}\, 
\frac{24 \, \alpha _1^2+44 \, \alpha _1\, \alpha _2+21 \, \alpha _2^2}{2\, \alpha_1 + \alpha_2}\, , \quad 
\delta^{(40)}  \mu_1 = - \frac{\tau_2^3}{45} \, 
\frac{4 \, \alpha _1^2+19 \, \alpha _1 \, \alpha _2+16 \, \alpha _2^2}{\left(\alpha _1+\alpha _2\right) 
\left(2 \, \alpha_1+\alpha _2\right)}
\end{equation}
$\delta^{(04)} z$ and $\delta^{(04)}  \mu_1$ with expressions
\begin{equation}
\frac{\delta^{(04)} z}{\delta^{(20)} z} = \frac{ 3\, \tau_2^2  }{5} \, 
\frac{\alpha_1+ 3\, \alpha_2}{2\, \alpha_1 + \alpha_2}\, , \quad 
\delta^{(04)}  \mu_1 = - \frac{\tau_2^3}{10} \, 
\frac{2\,\alpha_1^2 + 7\, \alpha_1\, \alpha_2+8\,\alpha_2^2}{\left(\alpha_1+ \alpha_2\right)\left(2\, \alpha_1 + \alpha_2\right)}
\end{equation}
$\delta^{(12)} \tau$, $\delta^{(12)} \mu_1$ and $\delta^{(12)} \mu_2$ with expressions
\begin{eqnarray}
&& \qquad \qquad \quad \quad 
\delta^{(12)} \tau = -\frac{\tau _2^4}{2}\, \frac{\alpha _2 \left(\alpha _1+\alpha _2\right)  \left[J\, \alpha_2 + 
2 \left(\alpha _1+\alpha _2\right)^2\right]}{\alpha _1 \left(2 \, \alpha _1+\alpha _2\right)^3}
\\[5pt]
&&
\delta^{(12)} \mu_1 = \frac{\tau _2^4}{2}\,  \frac{4 \, \alpha _1^3+12 \, \alpha _1^2\,  \alpha _2+4 \, \alpha _2^3+
\alpha _2^2 \left(J+ 12 \, \alpha _1\right)}{\left(\alpha_1+\alpha _2\right) \left(2 \, \alpha _1+\alpha _2\right){}^2} \, , 
\quad \delta^{(12)} \mu_2 = - \frac{\alpha _1+\alpha _2}{\alpha _1}\, \delta^{(12)} \mu_1
\nonumber
\end{eqnarray}
$\delta^{(32)} \tau$, $\delta^{(32)} \mu_1$ and $\delta^{(32)} \mu_2$ with expressions
\begin{eqnarray}
&&
\delta^{(32)} \tau = -\frac{\tau _2^6}{12}\, \frac{\alpha _2 \left(\alpha _1+\alpha _2\right) 
\Big[12 \, \alpha _1^3+20 \, \alpha _2^3+4 \, \alpha _1 \, \alpha _2 \left(11 \, \alpha _1+
2 \, J\right)+\alpha _2^2 \left(52 \, \alpha _1+11 \, J\right)\Big]}{\alpha _1 \left(2 \, \alpha _1+\alpha _2\right){}^4}
\nonumber \\[5pt]
&&
\delta^{(32)} \mu_1 = \frac{\tau _2^4}{12}\,  \frac{8 \, \alpha _1^4+44 \, \alpha _1^3 \,  \alpha _2 +
84 \, \alpha _1^2 \, \alpha _2^2+ 2 \, \alpha _1 \, \alpha _2^2 \left(34 \, \alpha _2+5 \, J\right) + 
\alpha _2^3 \left(20 \, \alpha _2+9 \, J\right)}{\left(\alpha _1+\alpha _2\right) \left(2 \, \alpha _1+\alpha _2\right){}^3} 
\nonumber \\[5pt]
&&  \qquad \qquad \quad \quad  \qquad \qquad  \qquad  \quad 
\delta^{(32)} \mu_2 = - \frac{\alpha _1+\alpha _2}{\alpha _1}\, \delta^{(32)} \mu_1
\end{eqnarray}
$\delta^{(24)} z$ with expression
\begin{eqnarray}
\frac{\delta^{(24)} z}{\delta^{(20)} z} &=&  \frac{ \tau_2^4 }{140}\, 
\frac{1}{\alpha_1^2\left(2\, \alpha_1 + \alpha_2\right)^5}\Bigg[2032 \, \alpha _1^7+12080 \, \alpha _2 \, \alpha _1^6+
10 \, \alpha _2^2 \, \alpha _1^3 \left(2251 \, \alpha _2^2+84 \, J^2+1218 \, \alpha _2 \, J\right) 
\nonumber \\[5pt]
&&+ \, 6 \alpha _2^3 \, \alpha _1^2 \left(1221 \, \alpha _2^2+280 \, J^2+1330 \, \alpha _2 \, J\right) + 
240 \, \alpha _2 \, \alpha _1^5 \left(111 \, \alpha _2+7 \, J\right) 
\nonumber \\[5pt]
&&+\,  40 \, \alpha _2^2 \, \alpha _1^4 \left(809 \, \alpha _2+189 \, J \right)+105 \, \alpha _2^4 \, \alpha _1 \, J 
\left(12 \, \alpha _2+7 \, J \right) - 105 \, \alpha _2^5 \left(2 \, \alpha_2 + J \right){}^2\Bigg]
\end{eqnarray}
and $\delta^{(24)} \mu_1$ with expression
\begin{eqnarray}
\delta^{(24)} \mu_1&=&  - \frac{ \tau_2^5 }{280}\, 
\frac{1}{\alpha_1 (\alpha_1+\alpha_2)\left(2\, \alpha_1 + \alpha_2\right)^4}\Bigg[528 \, \alpha _1^6 + 
3392 \, \alpha _2 \, \alpha _1^5+840 \, \alpha _2^6+8 \, \alpha _2^3 \, \alpha _1 \times 
\nonumber \\[5pt]
&& \left(1556 \, \alpha _1^2+35 \, J^2+315 \, \alpha _1 \, J\right) + \alpha_2^4 \left(9916 \, \alpha _1^2+ 245 \, J^2+
2520 \, \alpha _1 \, J\right) +8 \, \alpha _2^2 \, \alpha _1^3  \times 
\nonumber \\[5pt]
&& \left(1138 \, \alpha _1+105 \, J\right) + 24 \, \alpha _2^5 \left(184 \, \alpha _1+35 \, J\right)\Bigg] \, . 
\end{eqnarray}


\end{document}